\documentclass[twocolumn,pre,final,showpacs,floatfix]{revtex4-1}
\usepackage{amssymb,amsmath}
\usepackage{bm}
\usepackage{graphicx}
\usepackage{epstopdf}

\newcommand{\Tu} {$T_{\rm u}$}
\newcommand{\Tum} {T_{\rm u}}

\begin{document}

\title[Universal Berezinskii-Kosterlitz-Thouless dynamic scaling in the intermediate time range in frustrated Heisenberg antiferromagnets on triangular lattice ] 
{Universal Berezinskii-Kosterlitz-Thouless dynamic scaling in the intermediate time range in frustrated Heisenberg antiferromagnets on triangular lattice}

\author{Ivan S. Popov, Pavel V. Prudnikov}
\affiliation{Omsk State University,  pr. Mira 55, Omsk 644077 Russia} 

\author{Andrey N. Ignatenko}
\affiliation{Institute of Metal Physics, 620990, Kovalevskaya str. 18 Ekaterinburg, Russia}

\author{Andrey A. Katanin}
\affiliation{Institute of Metal Physics, 620990, Kovalevskaya str. 18 Ekaterinburg, Russia}
\affiliation{Ural Federal University, 620002, Mira str. 19, Ekaterinburg, Russia}

\begin{abstract}
We investigate non-equilibrium properties of the frustrated Heisenberg antiferromagnets on the triangular lattice. Nonequilibrium critical relaxation of frustrated Heisenberg antiferromagnets shows a dynamic 
transition (or, at least, sharp crossover) 
at the same temperature  $\Tum=0.282 J$ as for static properties due to unbinding of $\mathbb{Z}_2$-vortices. 
We show that starting from the high-temperature initial state, due to presence of $\mathbb{Z}_2$-vortices in the considering system, in a broad temperature range $T<\Tum$ the dynamic properties in the intermediate time range are similar to those of two-dimensional XY model below Berezinskii-Kosterlitz-Thouless transition. The interaction of $\mathbb{Z}_2$-vortices with spin-wave degrees of freedom does not emerge until rather long times.
\end{abstract}


\maketitle

\section{Introduction}

Investigations of critical behavior of 
systems with continuous symmetry of the order parameter attract a lot of attention and represent considerable fundamental and practical interest  \cite{Ma}.
Strong fluctuation effects playing an important role in low dimensional \cite{Auerbach,Book_de_Jongh_1990} and frustrated \cite{Book_frustr_Lacroix_Mendels_Mila} systems, may lead to non-trivial critical behavior of these systems.

In two-dimensional systems with continuous symmetry of the order parameter the long-range order at finite temperatures is destroyed by spin fluctuations \cite{Mermin}. Presence of topologically non-trivial configurations (vortices), however, leads to the peculiarities of the disordered (paramagnetic) state. In particular, a topological phase transition,  associated with the dissociation of vortex pairs, occurs at a temperature $T_{\rm BKT} \neq 0$, as was shown by Berezinskii \cite{Berezinskii_1,Berezinskii_3}, and then Kosterlitz and Thouless \cite{Kosterlitz_Thouless,Kosterlitz}. Below $T_{\rm BKT}$ vortices are bound in pairs, the correlation length is infinite, and there is an algebraic (power-law) decay of spin correlators. This changes at $T>T_{\rm BKT}$, when vortices are not bound, causing finite correlation length. 

The presence of vortices in non-equilibrium states (even below $T_{\rm BKT}$) may change essentially the dynamical properties of the system. In particular, in Ref. \cite{PhysRevLett.84.1503} it was shown that in the presence of vortices the time dependence of the correlation length $\xi(t)\propto (t/\ln t)^{1/2}$ acquires logarithmic correction. The dynamic spin correlation functions fulfill certain scaling relations, which involve the abovementioned time dependence of the correlation length.

Magnetic frustration, which occurs due to peculiarities of lattice geometry, yields on one hand an  enhancement of the fluctuation effects, but on the other hand, provides a possibility of realizing new types of topological structures with respect to those in non-frustrated systems.  
Experimental studies of 
magnetic materials with triangular lattice, in particular $\rm NaCrO_{2}$ \cite{Exp_1,Exp_2,Exp_3,Exp_4}, $\rm
NiGa_{2}S_{4}$ \cite{Exp_5,Exp_6,Exp_7,Exp_8, Exp_9,Exp_10,Exp_11},
$\rm \kappa-(BEDT-TTF)_{2}Cu_{2}(CN)_{3}$
\cite{Exp_12,Exp_13,Exp_14} и $\rm EtMe_{3}Sb[Pd(dmit)_{2}]_{2}$
\cite{Exp_15,Exp_16}, reveal distinct anomaly behavior at non-zero temperatures, typical of phase transitions and non-trivial dynamic properties above and below the observed phase transition \cite{Exp_1,Exp_4,Exp_7,Exp_9,Exp_10}.

Equilibrium properties of frustrated antiferromagnets on a triangular lattice exhibit features of the $\mathbb{Z}_2$-vortex unbinding  transition (or at least sharp crossover), which is similar to 2D Berezinskii-Kosterlitz-Thouless transition in XY-model \cite{Kawamura,Kawamura_exp}. 
The $\mathbb{Z}_{2}$-vortices (or their major part) are bounded to vortex pairs at temperatures below a certain characteristic temperature \Tu, which
allows to draw an analogy between the low temperature properties of
this system and critical properties of 2D XY model. In contrast to the XY model below $T_{\rm BKT}$, for frustrated magnetic materials
the correlation length is finite even below $\Tum$  
due to the contribution of non-topological (or spin-wave) degrees of freedom \cite{Azaria,OurNP,MC_Southern_Young_1993,Wintel_Everts_Apel,Kawamura1}. 

The finiteness of the correlation length and difference of $\mathbb{Z}_2$-vortices in respect to those in XY model lead to the question, to what extent the dynamic properties of frustrated systems are similar (or different) to those of the XY-model, in particular whether 
peculiar phenomena of the nonequilibrium critical dynamics, such as dynamic scaling and aging effects    \cite{PPP_JETPLett,2013_review,2013_Berthier,Berthier,Abiet}, occur in the low-temperature phase $T<\Tum$. Although the equilibrium dynamics of triangular lattice antiferromagnets was studied earlier \cite{Kawamura_dynamics},
in the present paper we concentrate on the nonequilibrium dynamic properties and show, that at intermediate timescales the dynamic properties of frustrated magnets are in fact very similar to those of XY model. 

\section{Models and approaches}

We consider the non-equilibrium dynamics of the frustrated Heisenberg model on a triangular lattice
\begin{equation}\label{H_S_orded}
    H=J \sum_{<i,j>}{\mathbf{S}_i \mathbf{S}_j}, 
\end{equation}
where $J>0$ is the exchange integral, which is taken as a unit of energy in the present study, $\mathbf{S}_i$ are classical Heisenberg spins located at sites $i$ of the two-dimensional triangular lattice, $<i,j>$ denotes summation over the nearest neighbors of the lattice. The ground state of the system has the form of 120-degree order with the wave vector $\mathbf{Q}=(4\pi/3,0)$.



In the continuum approach, the dynamics of the triangular lattice Heisenberg antiferromagnet can be described by fluctuating order parameter having the form of continuous vector fields $\mathbf{e}_{1}(\mathbf{x})$, $\mathbf{e}_{2}(\mathbf{x})$ defined by the equation
\begin{equation}
\label{eqn_low_energy_conf}
\mathbf{S}_{i}=\mathbf{e}_{1}(\mathbf{R}_{i})\cos(\mathbf{Q}\cdot \mathbf{R}_{i})+\mathbf{e}_{2}(\mathbf{R}_{i})\sin(\mathbf{Q}\cdot  \mathbf{R}_{i}).
\end{equation}
These fields evolve in time according to a stochastic equation (dynamic model A in Hohenberg and Halperin classication \cite{Hohenberg})
\begin{equation}
\label{Stoch} 
	\frac{\partial \mathbf{e}_{\alpha}({\bf x},t)}{\partial t}=-\Gamma \frac{\delta \mathcal{H}}{\delta \mathbf{e}_{\alpha}({\bf x},t)}+\gamma_{\alpha}({\bf x},t). 
\end{equation}
Here $\Gamma$ is the kinetic coefficient, $\gamma_{\alpha}({\bf x},t)$ is the random Gaussian force, $\alpha=1, 2$, and $\mathcal{H}$ is the continuum limit of the Hamiltonian (\ref{H_S_orded}), cf. Ref. \cite{DR,OurNP1},
\begin{equation}
\label{eqn_NLsigma_model_0}
\mathcal{H}=\frac{\rho_{\perp}}{2}\int d^2\mathbf{x}\left[(\nabla \mathbf{e}_{1})^{2}+(\nabla \mathbf{e}_{2})^{2} + V(\mathbf{e}_{1},\mathbf{e}_{2})\right], 
\end{equation}
where $\rho_{\perp}\approx\sqrt{3}J/4$ is the transverse spin stiffness. Since $\mathbf{S}_{i}^2=1$, Eq. (\ref{eqn_low_energy_conf}) implies $\mathbf{e}_{1}^{2}\approx\mathbf{e}_{2}^{2}\approx 1$, $\mathbf{e}_{1}\cdot \mathbf{e}_{2}\approx0$ at each lattice site ${\bf R}_i$, the potential $V(\mathbf{e}_{1},\mathbf{e}_{2})$ must have degenerate deep minima at these $\mathbf{e}_{1,2}$. For fluctuations having sufficiently low energies  potential $V$ is reduced to an irrelevant constant which we choose to be zero.

For a numerical study of the nonequilibrium relaxation, we perform 
Monte Carlo simulations of the model (\ref{H_S_orded}) using Metropolis algorithm.
Although equilibrium properties were intensively studied previously \cite{Kawamura,MC_Kawamura_Kikuchi_1993,MC_Southern_Young_1993, MC_Southern_Xu_1995, Wintel_Everts_Apel, MC_Caffarel_et_al_2001, MC_Kawamura_Yamamoto_2007, Kawamura1}, the investigation of the  non-equilibrium processes allows us to get information on the dynamic properties of the frustrated systems. Time is set in Monte Carlo steps per spin ($\mathrm{MCS/s}$), which defines the time interval during which each of $N=L^2$ spins have the ability to change its state.
We consider lattices with sizes $L\le360$. 

\begin{figure}[h]\center
	\includegraphics[width=0.235\textwidth]{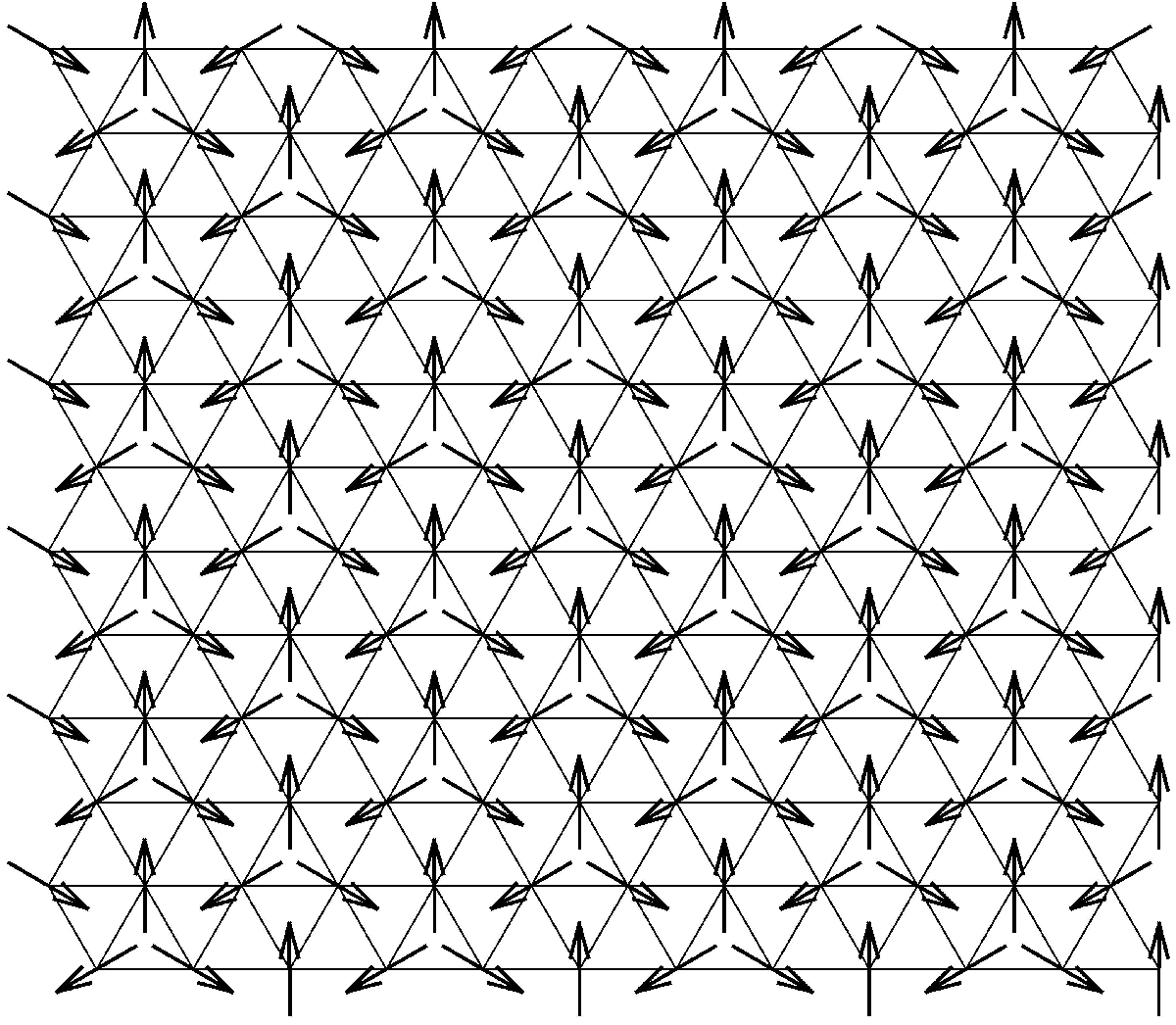}
	\includegraphics[width=0.235\textwidth]{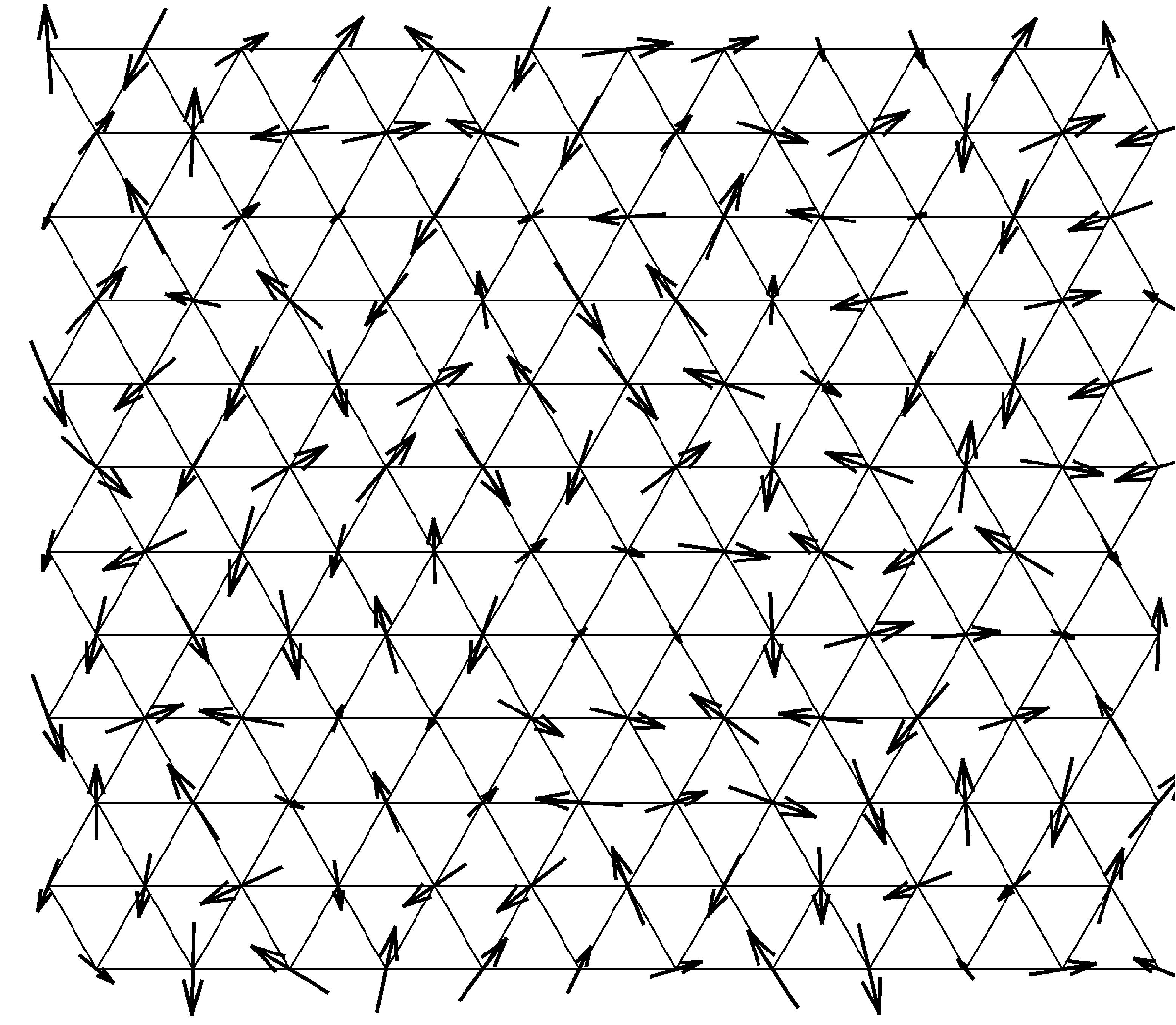}
	\caption{
Projection of spin vectors onto $xy$ plane for the initial low- and high temperature states of Monte Carlo simulation}
\label{Fig:Initial1}
\end{figure}

The low-temperature ordered state corresponds to the ground state of the system with the initial temperature $T^{\rm ini}_{L} = 0$. This initial state does not have vortex excitations. In contrast to this initial state, the high temperature initial state is prepared at temperature $T^{\rm ini}_{H}=20 \gg \Tum$ where the concentration of unbound vortices is much larger than at the equilibrium. Initial low- and high temperature states are visualized in Fig.~\ref{Fig:Initial1}.

The sequence of states defined by the Metropolis algorithm according 
to the transition probability between neighboring configurations, forms a Markov process.
The evolution of the non-equilibrium distribution function $P_n(t)$ 
can be written in the form of the master kinetic equation:
\begin{equation}
\label{master_eqn}
	\frac{dP_n(t)}{dt} = \sum_{m} \left[W(n \rightarrow m) P_n(t) - W(m \rightarrow n) P_m(t)\right],
\end{equation}	
where $W(n \rightarrow m) = \min \left[1.0, \exp(-\Delta E_{nm}/T)\right]$. 
The dynamics of the single-flip Metropolis algorithm corresponds to the dynamic model A, cf. Eq. (\ref{Stoch}). 

The time dependence of correlation length $\xi(t)$ can be defined as follows:
\begin{equation}
\xi(t) = \frac{3 L}{4 \pi} \sqrt{ \frac{\chi(t)}{\Phi(t)} - 1 },
\end{equation}
where 
\begin{equation}
\chi(t) = \left< \big|\mathbf{S}_Q(t)\big|^2 \right> - {\left<
\big|\mathbf{S}_Q(t)\big| \right> }^2
\end{equation}
is the susceptibility,
and 
\begin{equation}
\Phi(t) = \frac{1}{2}  \sum_{n=x,y} \left< \big| \sum_i
\mathbf{S}_i(t) e^{i \mathbf{q}_n\cdot\mathbf{R}_i + i \mathbf{Q}\cdot\mathbf{R}_i }\big|^2 \right>
\end{equation}
is the structural factor of system, 
$\mathbf{q}_{x} = (\frac{2\pi}{L}, -\frac{2\pi}{\sqrt{3}L})$,
$\mathbf{q}_{y} = (0, -\frac{4\pi}{\sqrt{3}L})$ are the 
vectors of the reciprocal lattice.




\section{$\mathbb{Z}_2$-vortex unbinding at equilibrium}
\label{sect_equilibrium}
Before proceeding to the nonequilibrium dynamics, let us consider first some important aspects of the $\mathbb{Z}_2$-vortex unbinding at equilibrium. 

Let there be an isolated vortex at the origin of coordinate system. If we neglect the spin-wave fluctuations (i.e. consider just a state with the lowest energy), the field configuration of the vortex,
\begin{equation}
\label{eqn_vortex_config}
\mathbf{e}_{1}+\mathrm{i}\, \mathbf{e}_{2}=e^{i\psi}\left(\mathbf{a}+\mathrm{i} \frac{\mathrm{x}\mathbf{b}+\mathrm{y}\mathbf{c}}{\sqrt{\mathrm{x}^2+\mathrm{y}^2}}\right),
\end{equation}
is parametrized by three real orthonormal vectors $\mathbf{a}$, $\mathbf{b}$, $\mathbf{c}$ and angle $\psi$ ($\mathrm{x}$, $\mathrm{y}$ are the Cartesian coordinates in the plane of the lattice). For general $\psi\ne \pi n/2$ ($n$ is an integer)  $\mathbb{Z}_2$-vortex is a non coplanar configuration. Nevertheless, the chirality vector 
\begin{equation}
\mathbf{e}_3=[\mathbf{e}_{1}\times \mathbf{e}_{2}]=\frac{-\mathrm{y}\mathbf{b}+\mathrm{x}\mathbf{c}}{\sqrt{\mathrm{x}^2+\mathrm{y}^2}}
\end{equation}
describes an ordinary vortex which is planar in spin space. The energy of the configuration (\ref{eqn_vortex_config})
\begin{equation}
\mathcal{H}\approx \pi\rho_{\perp}\ln(L/a)
\end{equation}
diverges logarithmically with the increase of the size of the system $L$ (here $a$ is the lattice spacing). Overall, if we neglect the spin-wave fluctuations, $\mathbb{Z}_2$-vortex is analogous to the vortex of the XY model (this analogy  extends to the interaction between $\mathbb{Z}_2$-vortices which is also logarithmic). 

However, this picture changes considerably when one accounts for the thermal spin-wave fluctuations. 
In contrast to the $XY$ model, the equilibrium correlation length at $T<\Tum$ is finite, although exponentially large at low temperatures $T\ll J$, where it is determined by long wavelength spin-wave fluctuations \cite{Azaria},
\begin{equation}
\label{xi_sw}
\xi_{\mathrm{sw}}(T)\propto a \exp{\frac{\pi(2+\pi)\rho_{\perp}}{T}}.
\end{equation}
These spin-wave fluctuations wash out the specific spin configuration of vortex at distances $r\gg\xi_{\mathrm{sw}}$ from the vortex core. 
As a result, the effective energy of an isolated vortex $E_{1}$, renormalized by the spin-wave fluctuations, does not diverge with an increase of the system size; at temperatures $T\ll \rho_{\perp}$ this energy for an infinite system ($L=\infty$) behaves as \cite{Ignatenko_tobepublished}
\begin{equation}
\label{eqn_E_1}
E_{1}\approx  \frac{3\pi^2\rho_{\perp}^2}{T}\propto \frac{T}{8}[\ln(\xi_{\mathrm{sw}}/a)]^2.
\end{equation}
Similarly, the two vortices almost cease to interact with each other at distances $r\gg \xi_{\mathrm{sw}}$, and their energy $E_{2}\approx 2E_{1}$. At shorter distances $r\lesssim\xi_{\mathrm{sw}}$ the double logarithmic attractive potential 
\begin{equation}
\label{eqn_vort_pot}
U(a\ll r)=2E_{1}-E_{2}\approx \frac{T}{4}[\ln(\xi_{\mathrm{sw}}/r)]^2
\end{equation}
emerges between vortices \cite{Ignatenko_tobepublished}.



The energy of the vortex (\ref{eqn_E_1}) is finite at $L=\infty$, so that finite density of unbound vortices $n_{\mathrm{u}}\sim a^{-2}\exp(-E_1/T)\propto \exp{(-\mathrm{const}/T^2)}$ will appear at equilibrium also below $\Tum$. Therefore, there are strictly speaking 
no differences in topological properties of the system at low and high temperatures, as well as no topological phase transition. Most likely this argument also means that there is a crossover rather than a 
phase transition (defined as non-analyticity of the free energy) at $T=\Tum$. However, the density of unbound vortices $n_{\mathrm{u}}(T)$ is extremely small at $T\le \Tum$ (according to our estimate, $n_{\mathrm{u}}(\Tum)\sim 10^{-23}$). Hence, in Monte Carlo simulations of finite systems with linear sizes $L\sim 100$, unbound vortices are not detected at $T<\Tum$ and $\mathbb{Z}_2$-vortex unbinding looks like a true topological phase transition, accompanied by 
peculiarities 
in free energy and its derivatives. The effect of unbound vortices below $\Tum$ for larger systems is also expected to be very small. 

As we argue in Sect. \ref{subsect_continual_noneq}, the correlation length (\ref{xi_sw}) provides an additional length scale not only for the equilibrium but also for the dynamics. 

\section{Non-equilibrium dynamic and static properties}

\subsection{Dynamics from low-temperature initial state towards equilibrium}

\begin{figure}[t!]
	\includegraphics[width=0.45\textwidth]{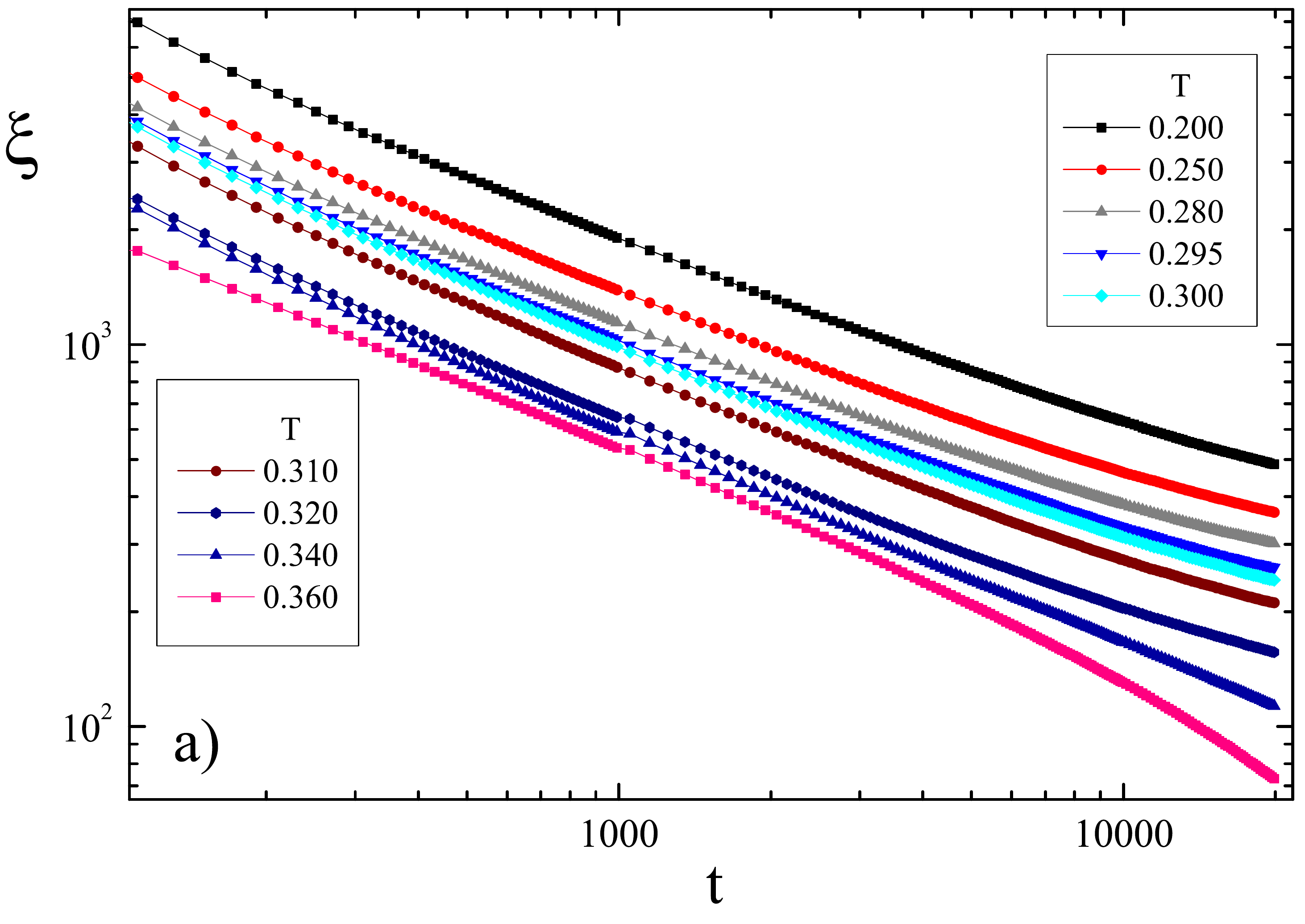}
	\includegraphics[width=0.45\textwidth]{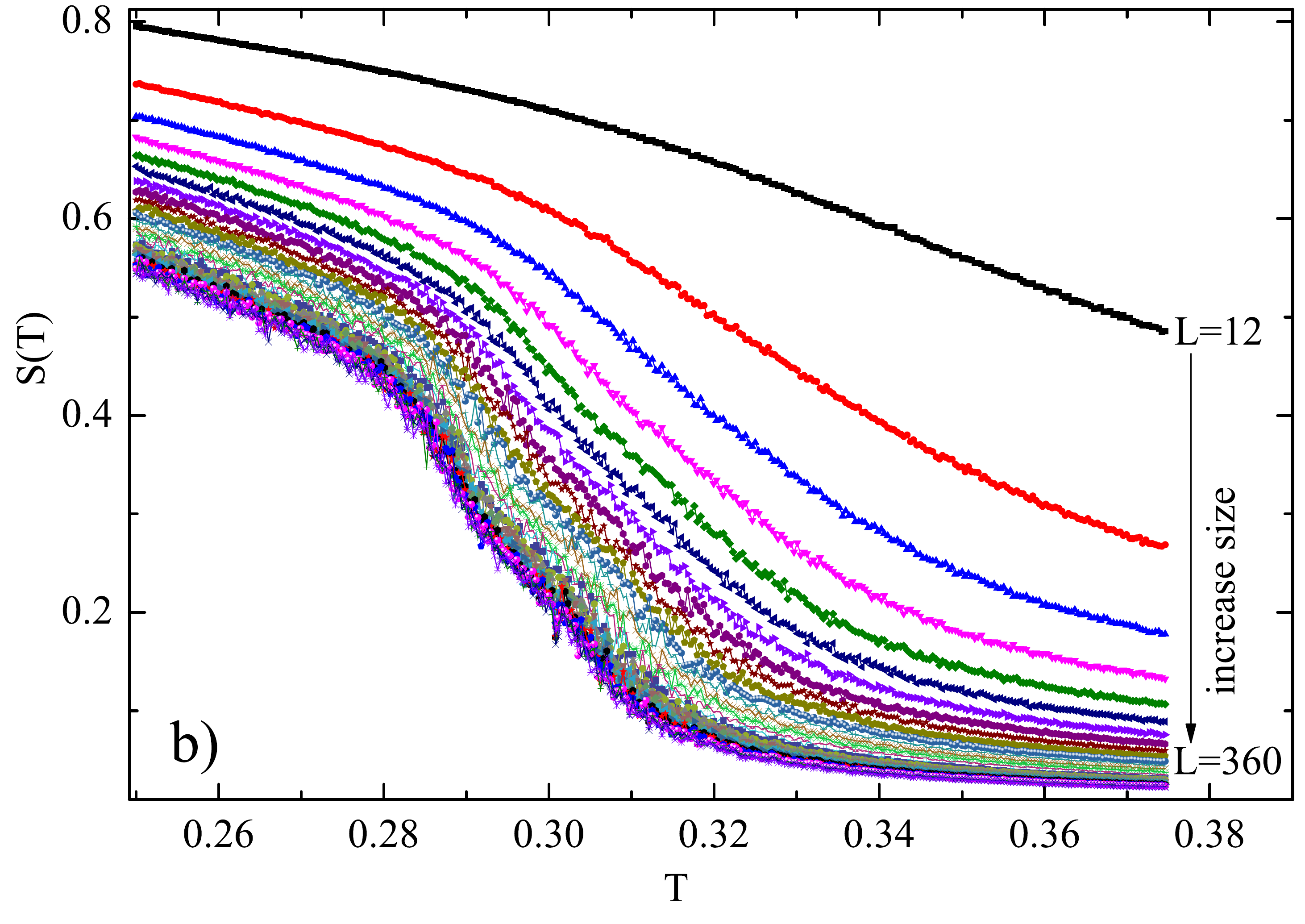}
\caption{\label{CL_1} (Color online) (a) The time dependence of the correlation length of the system 
    at nonequilibrium evolution from the low-temperature 
    initial state for the linear size $L=240$. (b) The temperature dependence of the equilibrium order parameter $S(T)$ for various sizes of the system $L$.
        }
\end{figure}

Starting from the low-temperature initial state, 
which is characterized by an infinitely large correlation length, 
the correlation length in the process of relaxation drops to its equilibrium value determined by the temperature and the linear size of the system (see Fig. \ref{CL_1}a).
One can see the existence of a certain characteristic $\mathbb{Z}_2$-vortex unbinding temperature
$\Tum\approx 0.3$, at which the correlation length at short time scales drops substantially, yielding also shorter correlation length in the limit of long times.

The temperature $\Tum$ in equilibrium can be accurately determined from the temperature dependence of the equilibrium order parameter 
$S(T)=\sqrt{|\mathbf{S}_{\bf Q}|^2}$, where
$
\mathbf{S}_{\bf Q} = ({1}/{N}) \sum_{i} {e^{i
\mathbf{Q}\mathbf{R}_i }\mathbf{S}_i }
$, 
which decreases with temperature, dropping at some (size-dependent) characteristic temperature $T^*(L)$ (see Fig. \ref{CL_1}b). 
The dependence of the position of the inflection points $T^*(L)$  of the temperature dependence of the order parameters
determines $\Tum$ and (quasi) critical exponent of the temperature dependence 
of correlation length of vortex subsystem $\xi_v \propto \exp(A/(T-\Tum)^\nu)$ according to
\begin{equation}
T^*(L)=\Tum+A/(\ln L)^{1/\nu}.
\label{TL}
\end{equation}
From fitting the least squares, we find $\Tum=0.282(5)$ and $\nu=0.37(5)$. The temperature $\Tum$ agrees well and the index $\nu$ agrees qualitatively with previous results $\Tum=0.285(5)$, $\nu=0.42\pm 0.15$ of Ref. \cite{Kawamura1}.

To analyze the dynamic behavior of magnetic order, in addition to $\chi(t)$ we have 
investigated the time dependence of the short-range order parameter $K(t)$
\begin{equation}
K(t) = \frac{1}{2 N} \left< \sum\limits_{\mathbf{r}} \big|
\mathbf{k}(\mathbf{r},t) \big| \right>,
\end{equation}
determined by local field of vector chirality $\mathbf{k}(\mathbf{r},t)$
\begin{equation}
\mathbf{k}(\mathbf{r},t) = \frac{2}{3\sqrt{3}} \left( \mathbf{S}_1
\times \mathbf{S}_2 + \mathbf{S}_2 \times \mathbf{S}_3 +
\mathbf{S}_3 \times \mathbf{S}_1 \right),
\end{equation}
where the summation is meant for all the elementary triangles of the lattice and spins with numbering $1$, $2$, and $3$ correspond to anti-clockwise circumvent of triangles.

\begin{figure}[t]
    \includegraphics[width=0.43\textwidth]{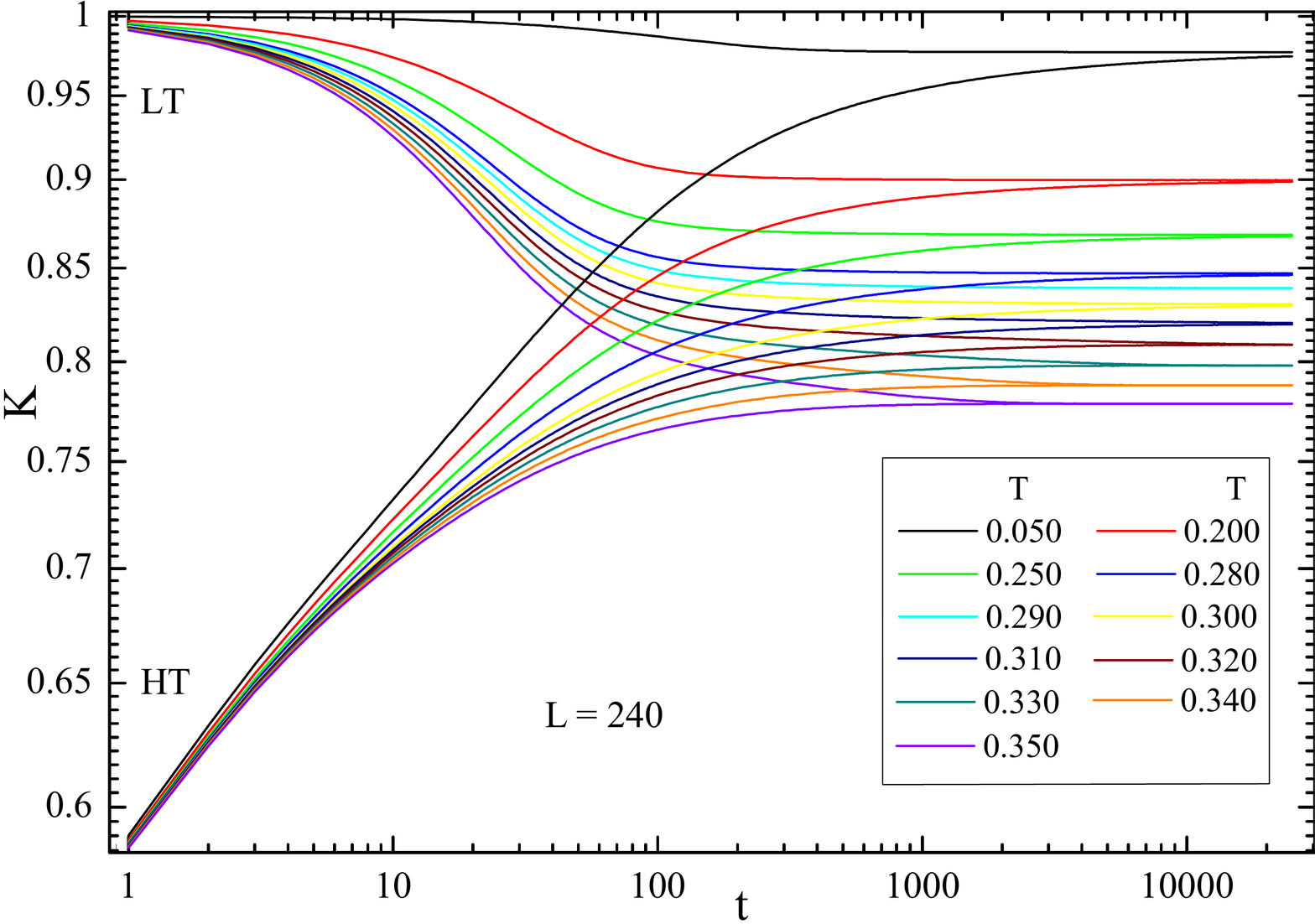}\vspace{-0.1cm}\\ 	
	\includegraphics[width=0.49\textwidth]{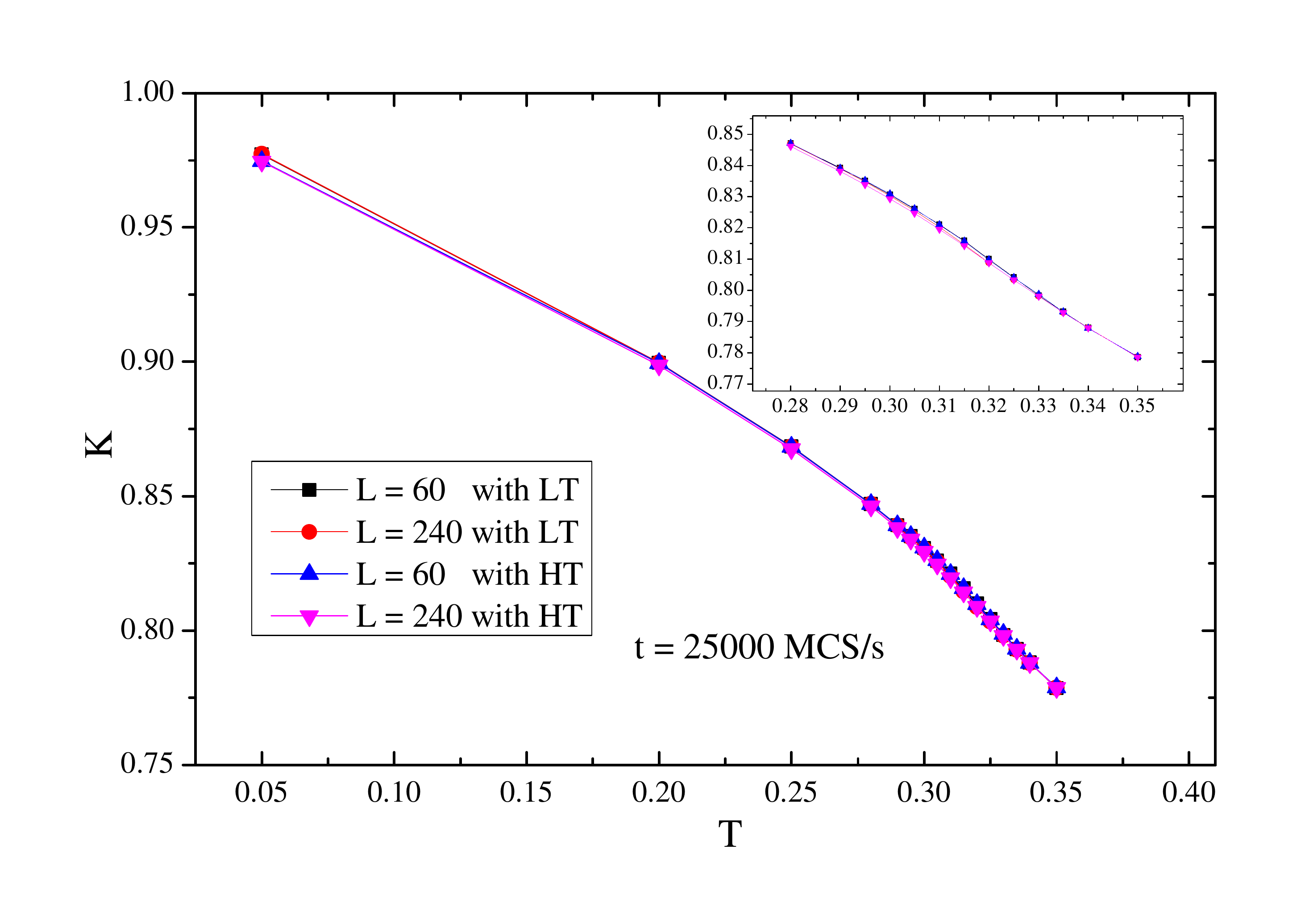}	
    \vspace{-0.5cm}
	\caption{\label{KT} (Color online) Upper plot: 
    the short range order parameter starting from the low-temperature (LT) and high-temperature (HT) initial state at L=240. Lower plot: temperature dependence of the equilibrium short-range order parameter for various $L$. The inset shows the behavior in the vicinity of $\mathbb{Z}_2$-vortex unbinding transition (or crossover).}
\end{figure}

Fig.~\ref{KT} shows the obtained time dependence $K(t)$ at various temperatures and temperature dependence of the equilibrium short-range 
order parameter of the system $K_T=K(t\rightarrow \infty)$. One can observe that the characteristic time of the onset of the equilibrium short-range order parameter $t_{\rm SRO}$ increases with decreasing temperature, and agrees with the time scale, starting from which the scaling of the time dependence of correlation length is observed, starting from the high-temperature initial state.
The equilibrium short-range order parameter $K_T$ monotonously decreases with temperature, and does not have any peculiarity (except possible inflection point) near $\Tum=0.282$.


\subsection{Dynamics from high-temperature initial state}

\subsubsection{Results of the continuum approach}
\label{subsect_continual_noneq}

Here we generalize the analysis of the vortex dynamics performed by Bray et. al for 2D XY-model  \cite{PhysRevLett.84.1503} to the case of the triangular lattice Heisenberg antiferromagnet. We use the continuum model (\ref{Stoch}) and the results of Sect. \ref{sect_equilibrium}.

\begin{figure}[t]
\includegraphics[width=0.46\textwidth]{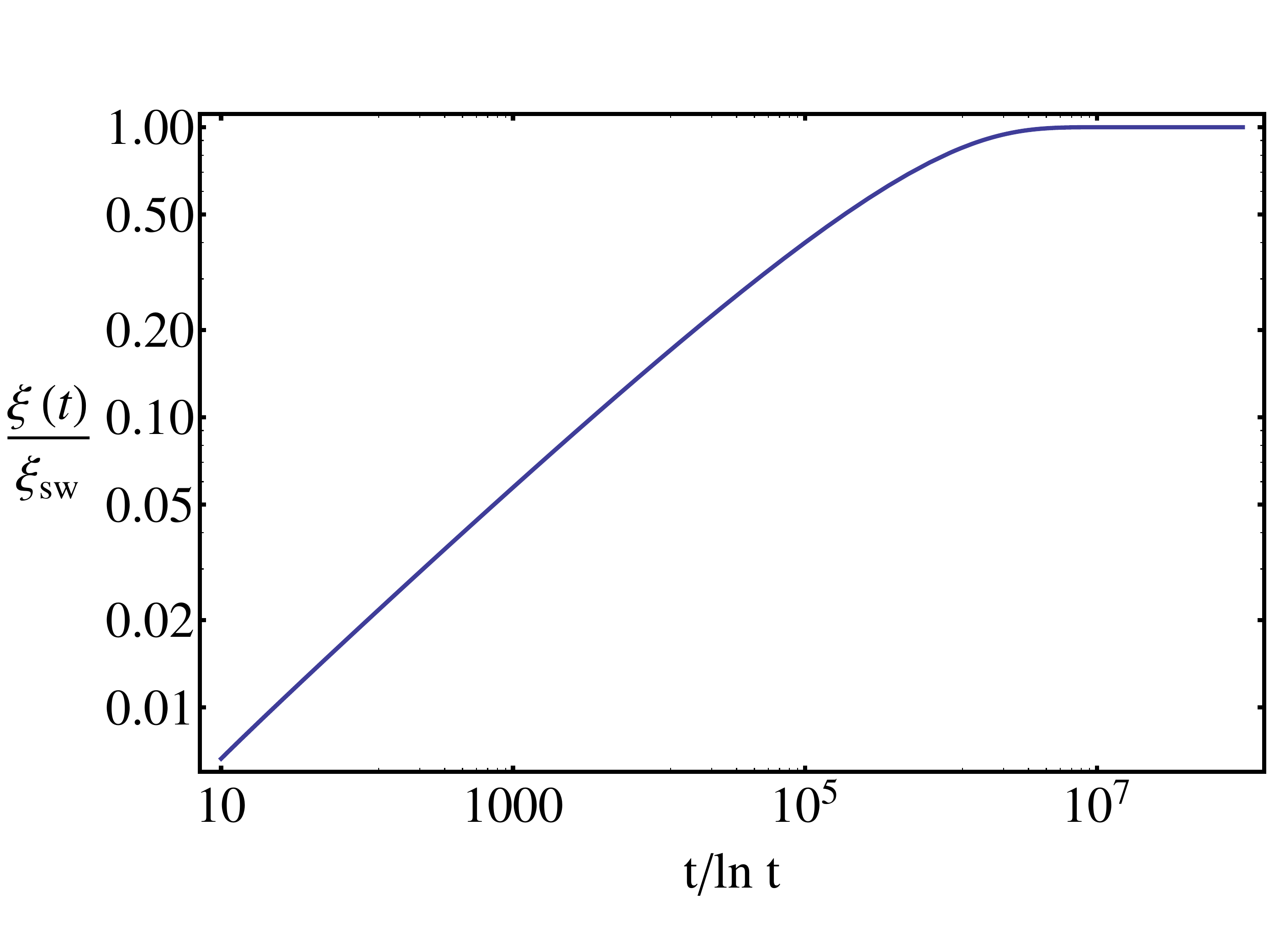}
\caption{\label{fig_xi_t_theory}Log-log plot of the function $\xi(t)$ satisfying equation (\ref{eqn_xi_dyn}), $t$ is measured in units $a^2/\Gamma$.}
\end{figure} 

Let us consider the relaxation of the high-temperature state. Disordered initial state contains many unbound vortices that are attracted to each other and annihilate as relaxation proceeds (see Fig.  \ref{Fig:Spindynamics} below). Against the background of this slow movement there is also more rapid spin wave dynamics, which is not related to a change in positions of vortex cores. Because of difference of characteristic times of the vortices and spin waves, the coarse-grained description of the relaxation process on basis of the adiabatic approximation becomes possible. 

Assuming that the spin-wave subsystem is already in the local (intermediate) thermodynamic equilibrium for each current position of the vortex cores, we introduce, following Ref. \cite{PhysRevLett.84.1503}, the friction coefficient $\gamma(r)$ which depends on the distance between vortices. To calculate $\gamma(r)$ we consider the field configuration $\mathbf{e}_{\alpha}(\mathrm{x},\mathrm{y},t)=\mathbf{e}_{\alpha}(\mathrm{x}-v t,\mathrm{y},t)$ corresponding to two vortices augmented by arbitrary spin wave fluctuation and moving as a whole with the velocity $v$ in direction $\mathrm{x}$. Then the rate of energy dissipation is $d\mathcal{H}/dt=\int d^{2}\mathbf{x}(\delta \mathcal{H}/\delta \mathbf{e}_{\alpha})\cdot (\partial \mathbf{e}_{\alpha}/\partial t)\approx-(1/\Gamma)\int d^{2}\mathbf{x}\sum_{\alpha}(\partial \mathbf{e}_{\alpha}/\partial t)^2=-\mathcal{H} v^2/\Gamma$ (we have neglected the stochastic term in equation (\ref{Stoch}), see Ref. \cite{PhysRevLett.84.1503}). Hence, the friction coefficient equals $\gamma(r)=\mathcal{H}/\Gamma$. Averaging over the spin-wave fluctuations using Eqs. (\ref{eqn_E_1}) and (\ref{eqn_vort_pot}) we obtain 
\begin{equation}
\gamma(r)=E_2/\Gamma\approx\frac{T}{4\Gamma}\left([\ln (\xi_{\mathrm{sw}}/a)]^2 - [\ln (\xi_{\mathrm{sw}}/r)]^2 \right).
\end{equation}
Writing down the condition for the balance of the conservative $\mathbf{F}=-\nabla U(r)$ and the friction $-\gamma(r)d\mathbf{r}/dt$ forces we come to the equation
\begin{eqnarray}
\label{eqn_f}
\frac{dr}{dt}&=&-f(r),\\
f(r)&=&\frac{2\Gamma}{r}\frac{\ln(\xi_{\mathrm{sw}}/r)}{[\ln( \xi_{\mathrm{sw}}/a)]^2 - [\ln( \xi_{\mathrm{sw}}/r)]^2}, \quad a\ll r \lesssim\xi_{\mathrm{sw}} \notag
\end{eqnarray} 
(note that $f(r)>0$ everywhere in the domain of applicability). According to this equation the fall of $\mathbb{Z}_2$-vortices at each other occurs in a finite time $\tau(r_{0})$, which depends on initial distance between vortices $r_{0}$. The inverse function $\xi(t)=\tau^{-1}(t)$ determines time dependence of some characteristic length which satisfies the equation ${d\xi}/{dt}=f(\xi)$, or, in the explicit form,
%
\begin{equation}
\label{eqn_xi_dyn}
\frac{d\xi^2}{dt}=\frac{4\Gamma\ln(\xi_{\mathrm{sw}}/\xi)}{\ln( \xi/a)\left[2 \ln(\xi_{sw}/\xi) +\ln (\xi/a) \right]},
\end{equation}
and can be identified with time dependence of the  dynamical correlation length $\xi(t)$ for $a\ll \xi(t) \lesssim\xi_{\mathrm{sw}}$. 
Even for $T\sim \Tum$ the spin-wave correlation length $\xi_{\mathrm{sw}}$ has a very large value of order of several thousand lattice spacings \cite{MC_Southern_Young_1993, Wintel_Everts_Apel
}, and typically $\xi(t)\ll \xi_{\mathrm{sw}}$ in the Monte Carlo calculations. In this regime $f(\xi)\approx {\Gamma}/({\xi\ln[\xi/a]})$ has the same form as in the XY model \cite{PhysRevLett.84.1503}. Accordingly, the solution of equation (\ref{eqn_xi_dyn}) gives the standard behavior of the correlation length $\xi(t)\propto (t/\ln t)^{1/2}$. On the triangular lattice this regime exists up to rather long times $t\ll(a^2/4\Gamma)(\xi_{\mathrm{sw}}/a)^{4/3}$. At even longer times, when $\xi(t)$ increases to the value of order $\xi_{\mathrm{sw}}$ the growth rate is slowing down. At $\xi=\xi_{sw}$ the equation (\ref{eqn_xi_dyn}) gives $f(\xi)=0$. Correspondingly, the dependence of $\xi(t)$ saturates (see Fig. \ref{fig_xi_t_theory}). This should happen physically, because the equilibrium correlation length on a triangular lattice at $T>0$ is finite.

\begin{figure}[b!]
	\includegraphics[width=0.235\textwidth]{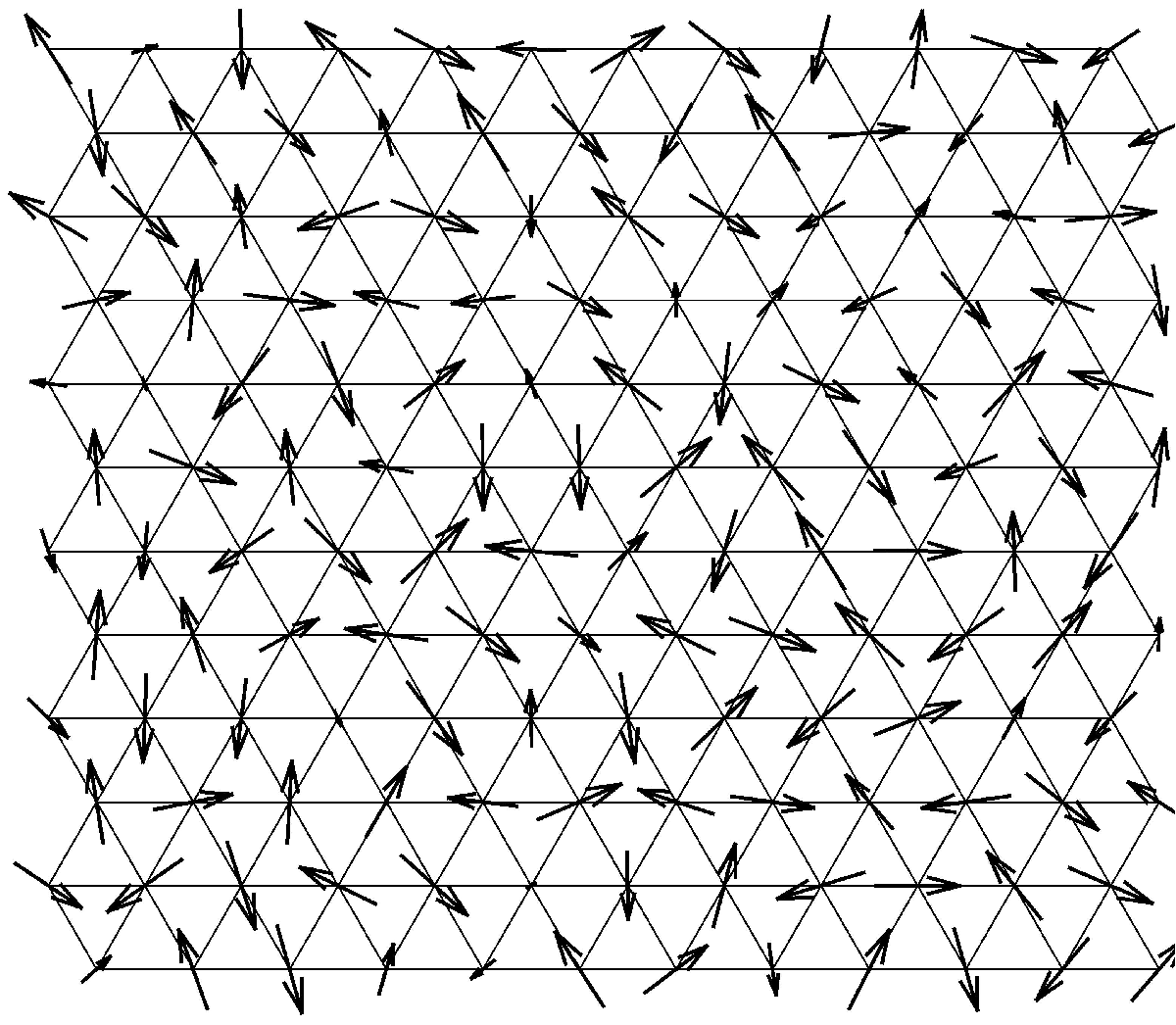}
	\includegraphics[width=0.235\textwidth]{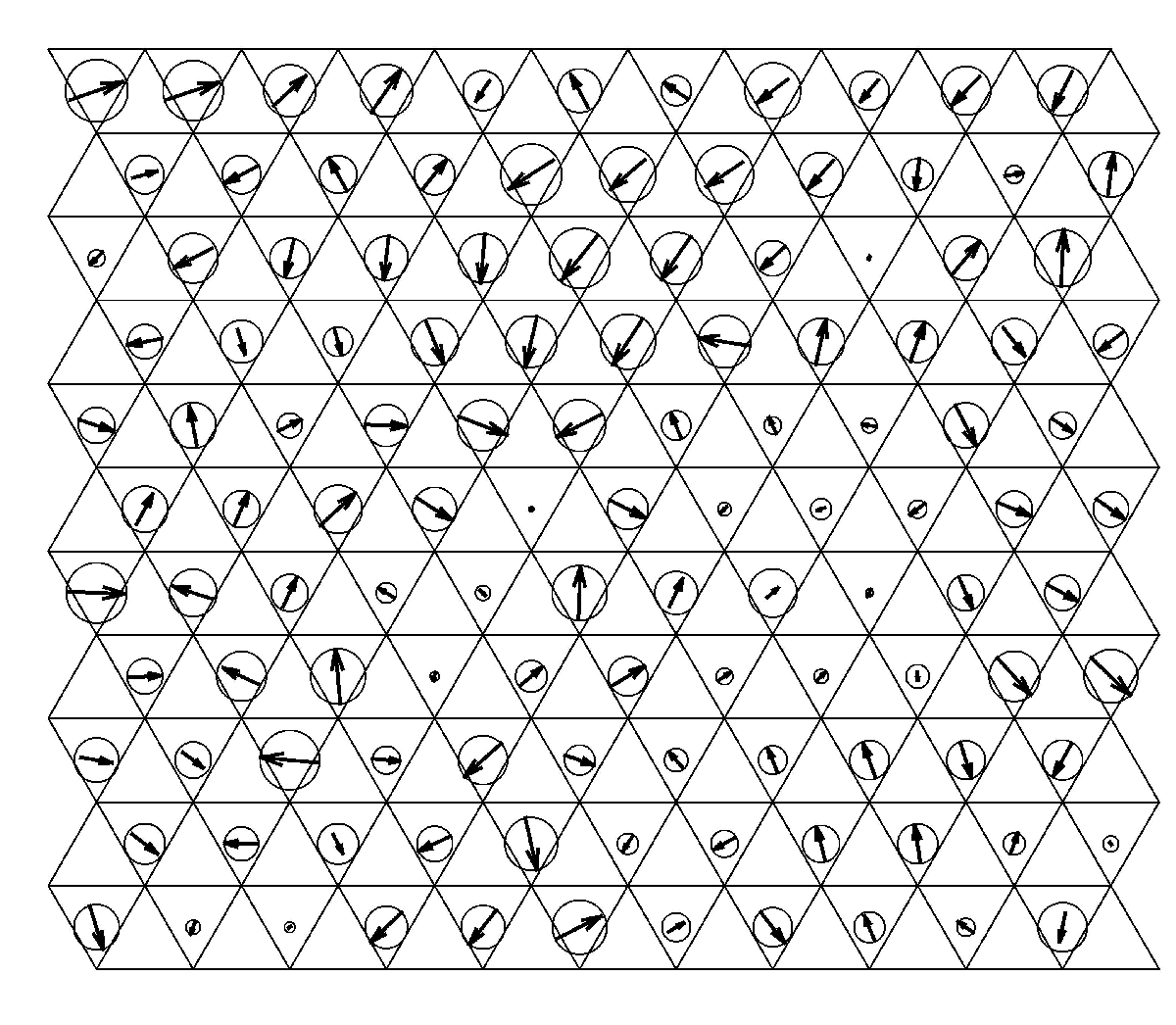}\\    	
                      \includegraphics[width=0.235\textwidth]{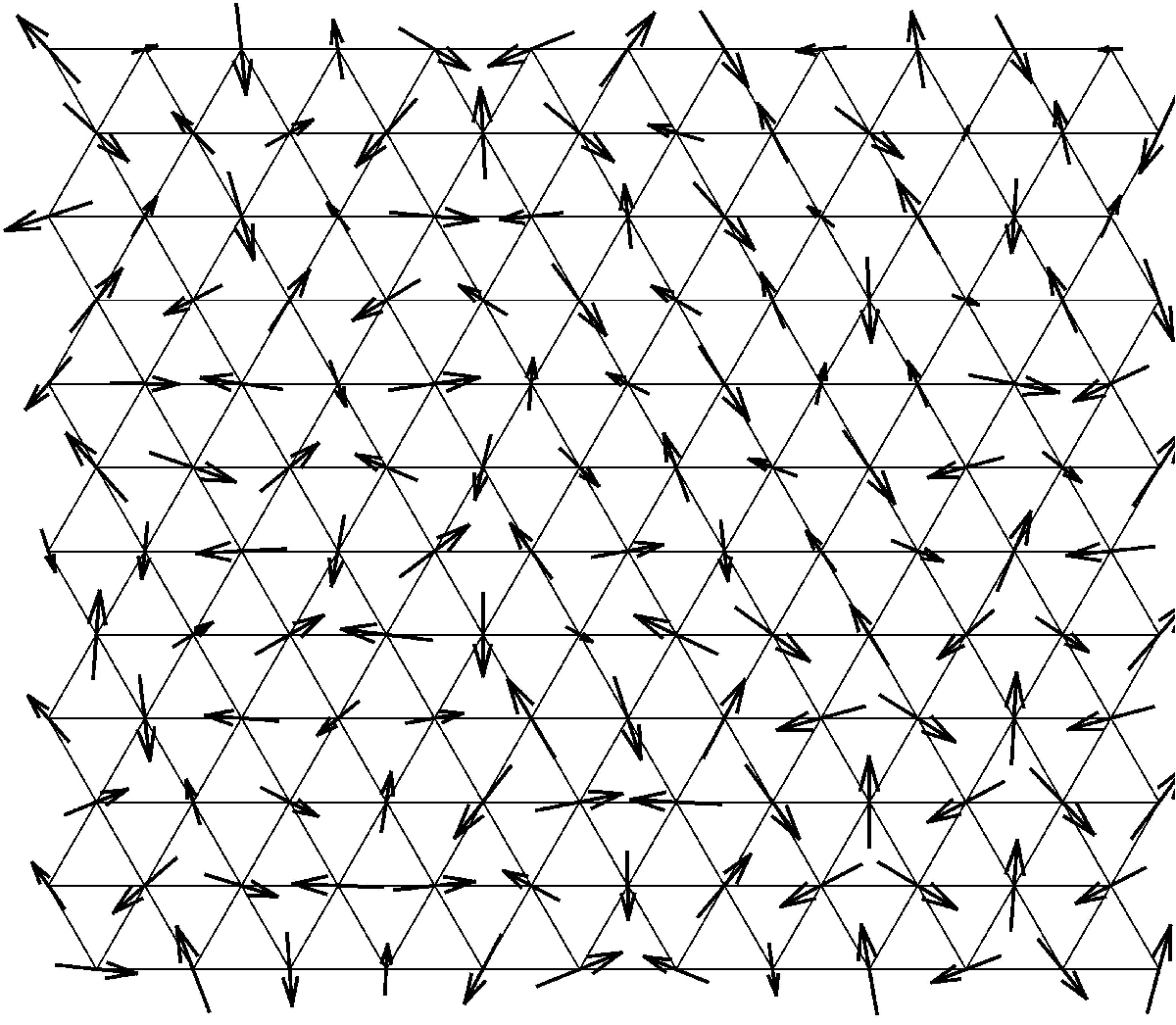}
	\includegraphics[width=0.235\textwidth]{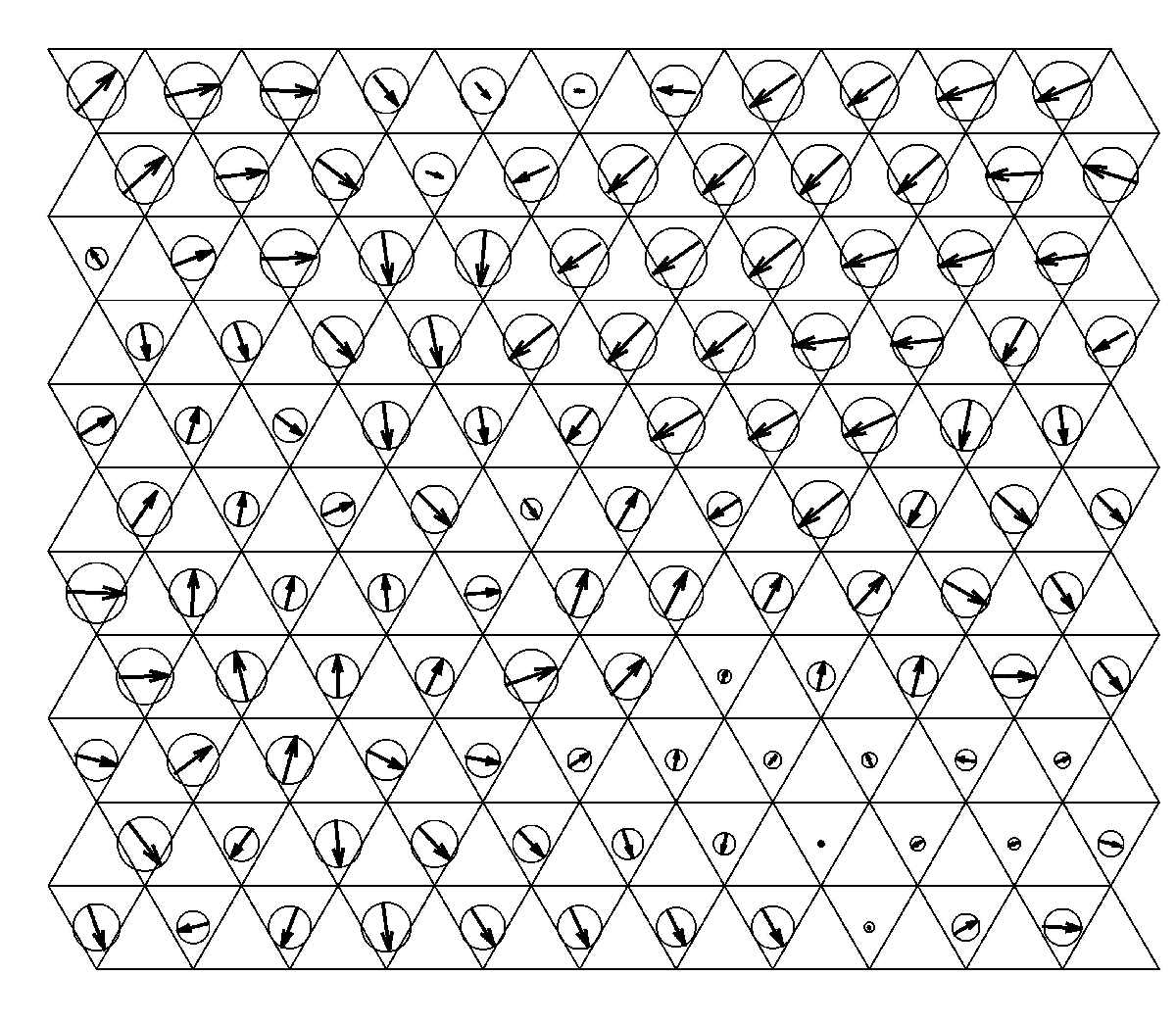}\\
    \includegraphics[width=0.235\textwidth]{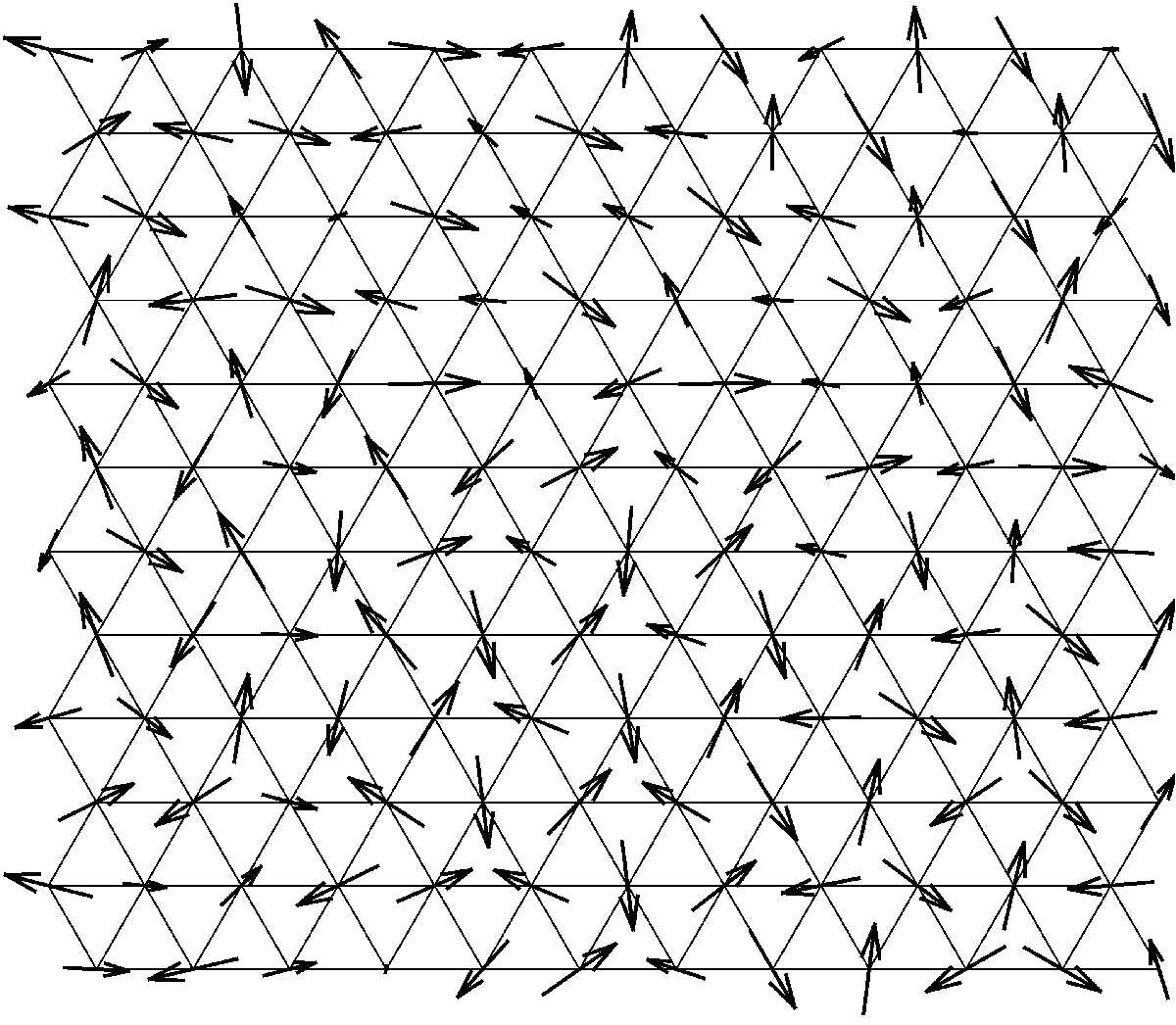}
    \includegraphics[width=0.235\textwidth]{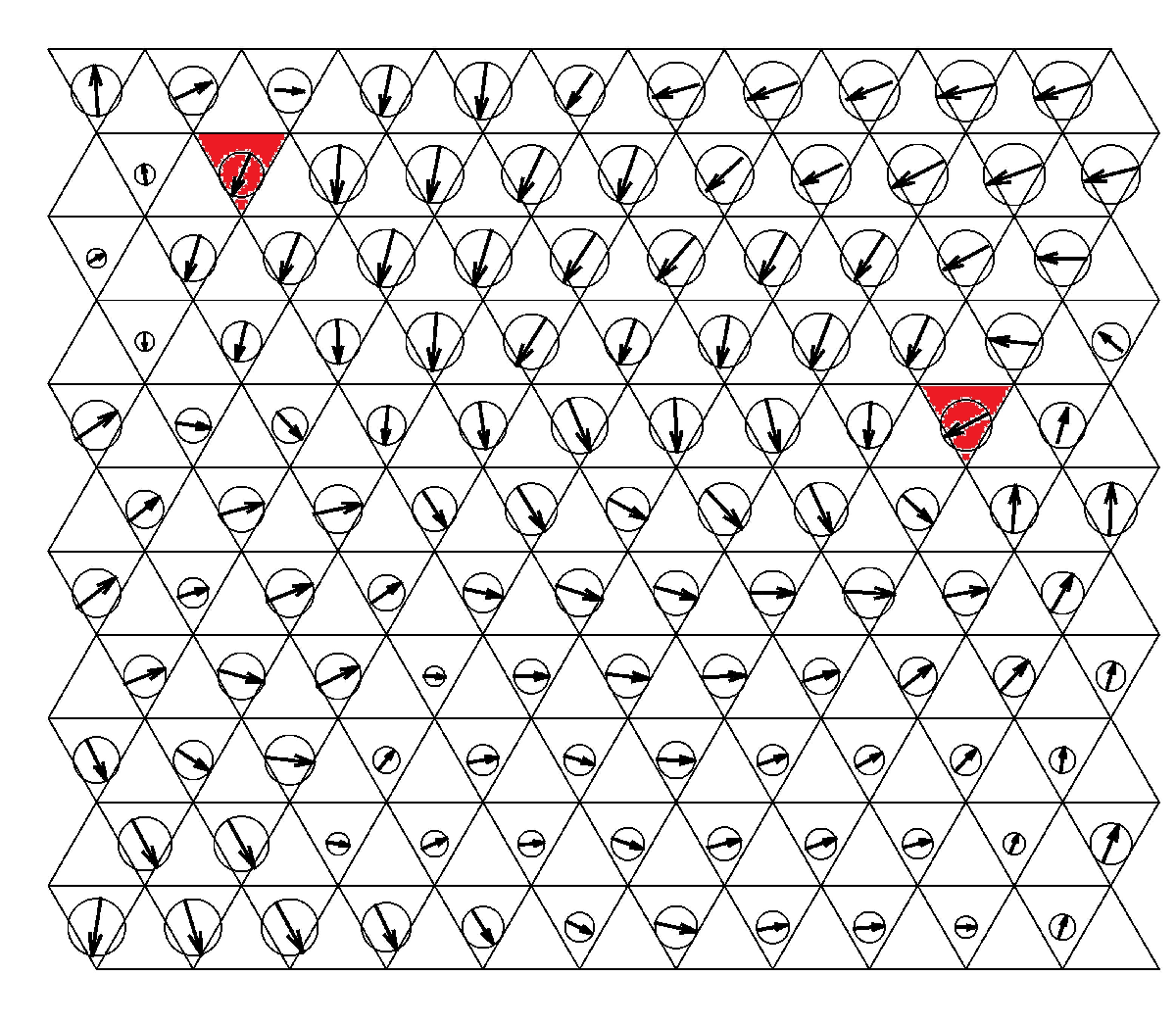}
	\caption{\label{Visual_Dynamics}
(Color online) Snapshot of relaxation dynamics of system at $T=0.1$ and the times $100$ (upper plots), 5000 (middle plots), and 15000 (lower plots) $\mathrm{MCS/s}$ from the high temperature initial state. Left (right) figures -- projections of the spins (chirality) vectors onto the $xy$-plane. The radius of each circle on right figures represents the length of corresponding chirality vector. Red triangles on the right lower figure denote cores of $\mathbb{Z}_{2}$-vortices; the vortices on the upper and middle figure are not shown: because of their large number, it is impossible to identify them uniquely.}
\label{Fig:Spindynamics}
\end{figure}

\subsubsection{Monte Carlo results}

\begin{figure}[t]
\vspace{-0.4cm}
    	\includegraphics[width=0.48\textwidth]{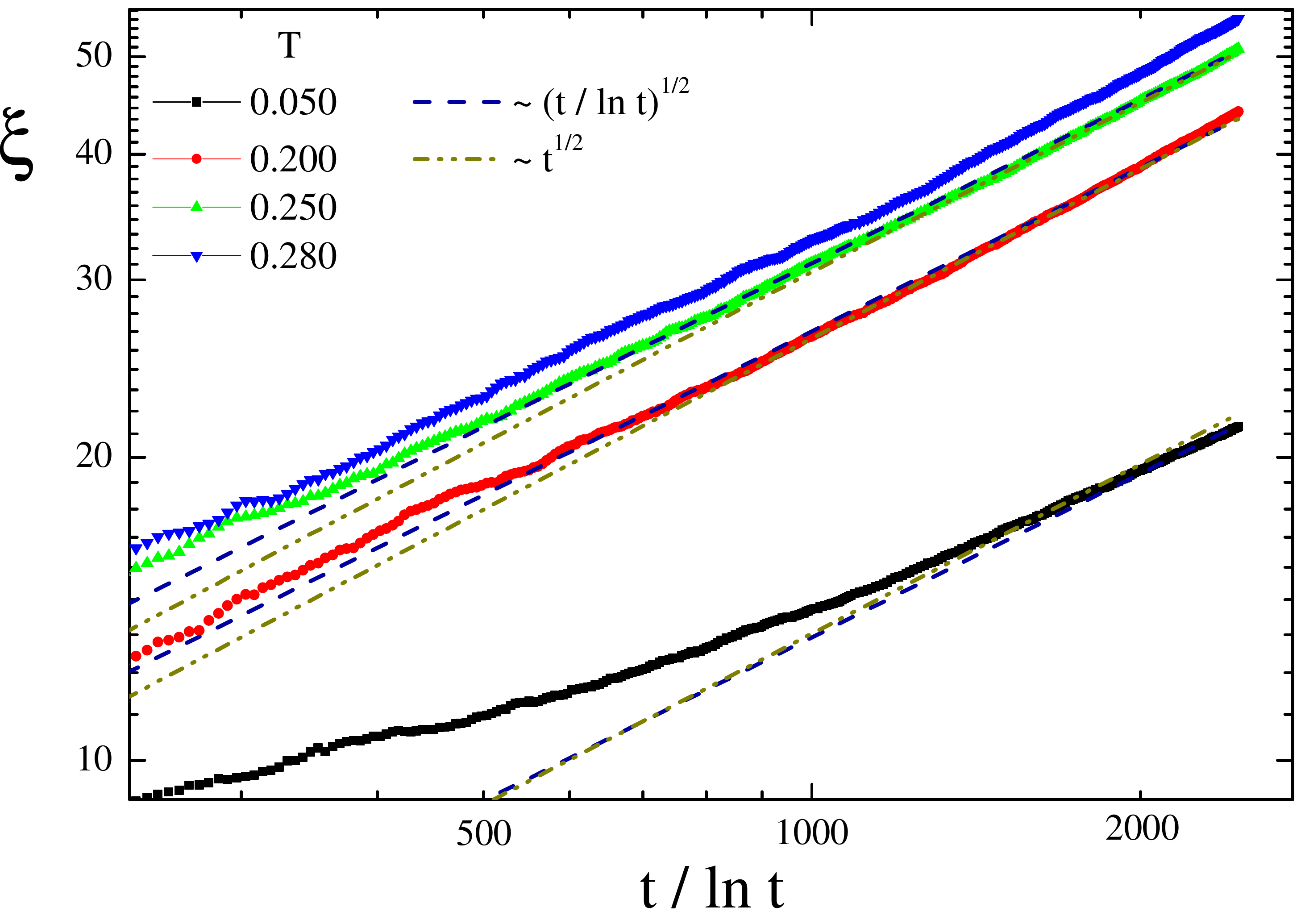}		
    \vspace{-0.6cm} 
	\caption{\label{ST} (Color online) The time dependence of the correlation length of the system 
    at nonequilibrium evolution from the high-temperature 
    initial state for the linear size $L=240$ with $(t/\ln t)^{1/2}$ (dashed lines) and $t^{1/2}$ (dot-dashed lines) fits. }
\end{figure}

On Fig. \ref{Fig:Spindynamics} we present the snapshots of spin configurations at different time (Monte Carlo steps), while on  Fig.~\ref{ST} we present the time dependence of the correlation length of the system from the initial high-temperature state, obtained from the Monte Carlo calculation, and plot as a function of $t / \ln t$ in a logarithmic scale. In agreement with the analytical consideration of previous subsection, the obtained time dependence of the correlation length is best fit by $(t/\ln t)^{1/2}$ law in the intermediate time range, 
reflecting violation of dynamic scaling due to the dynamics of vortex pairs, which are present in the system.

In Monte Carlo results of Fig. \ref {ST} we also observe a deviation from the universal dependence at low temperatures $T\le 0.2$ 
and short time scale $\le 100-1000$ $\mathrm{MCS/s}$. The time of the onset of scaling behavior increases with decreasing temperature and agrees with the time of the onset of strong short-range order $t_{\rm SRO}.$ 
Therefore, the observed deviation from $(t/\ln t)^{1/2}$ scaling behavior corresponds to the not fully formed short range magnetic order (see upper part of Fig. \ref{Fig:Spindynamics} for typical spin configuration), when vortices and spin waves are not yet well-defined. We emphasize, that the obtained time dependence of the correlation length in the intermediate time range is entirely due to the presence of $\mathbb{Z}_2$-vortices. 

\begin{figure}[t!]
\vspace{-0.5cm}
   	\includegraphics[width=0.48\textwidth]{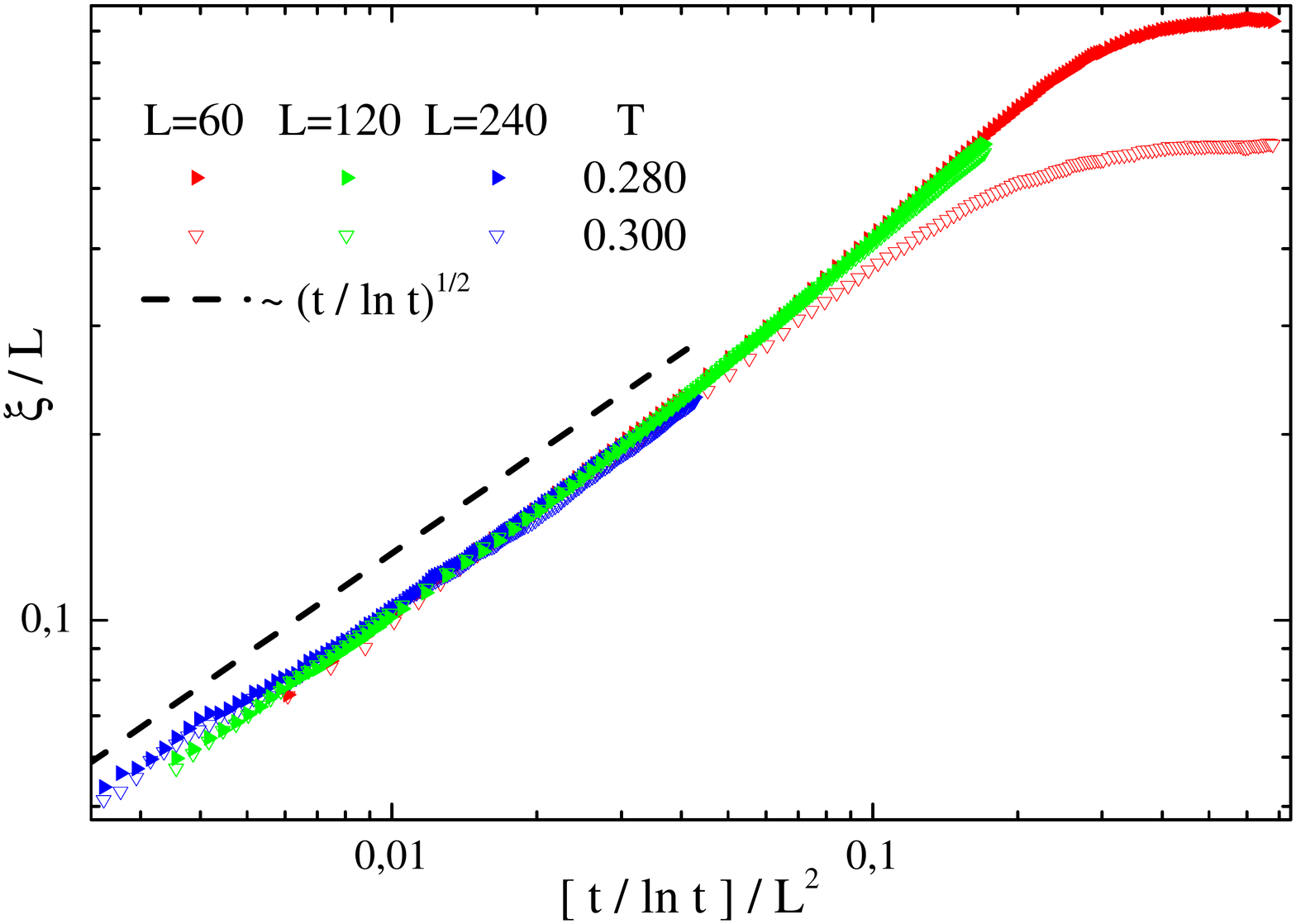}	
    \includegraphics[width=0.46\textwidth]{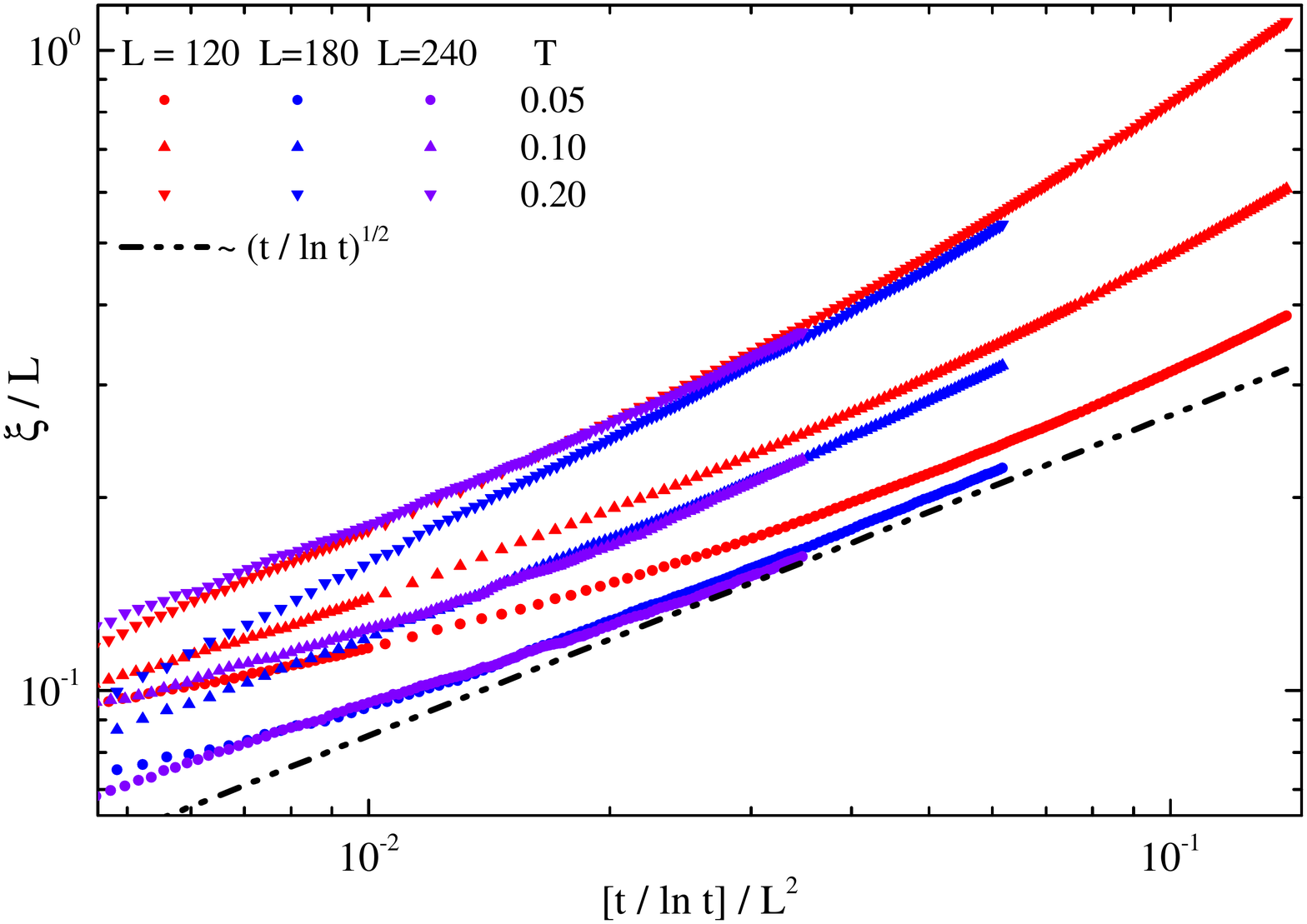}
    \vspace{-0.2cm}
 	\caption{\label{xsi} (Color online) The time dependence of correlation length $\xi(t)$ for different system sizes and high-temperature initial state, rescaled according to the dynamic exponent $z=2$ for triangular lattice (upper plot) and square lattice (lower plot). Dashed and dash-dotted lines show the slope of $(t/\ln t)^{1/2}$ fit.}
\end{figure}


To emphasize the scaling of the correlation length with the size of the system, we plot in Fig. \ref{xsi} the dependencies $\xi (t)$ for different system sizes, rescaled according to the dynamic exponent $z=2$. One can see that the dynamic scaling of $\xi(t)/L=f(t/(L^2 \ln t))$ is perfectly fulfilled with the function $f(x)\propto x^{1/2}$ at small $x$, implying in the intermediate time range $\xi \propto (t/\ln t)^{1/2}$ independently of the size $L$ at sufficiently large $L$. Note that similar dependence of $\xi(t)/L$ as a function of $t/L^2$ does not show universal scaling behavior, reflecting again violation of the "naive" dynamic scaling behavior.

\begin{figure*}[t!]
\begin{minipage}[t]{0.46\textwidth}
\includegraphics[width=0.9\linewidth]{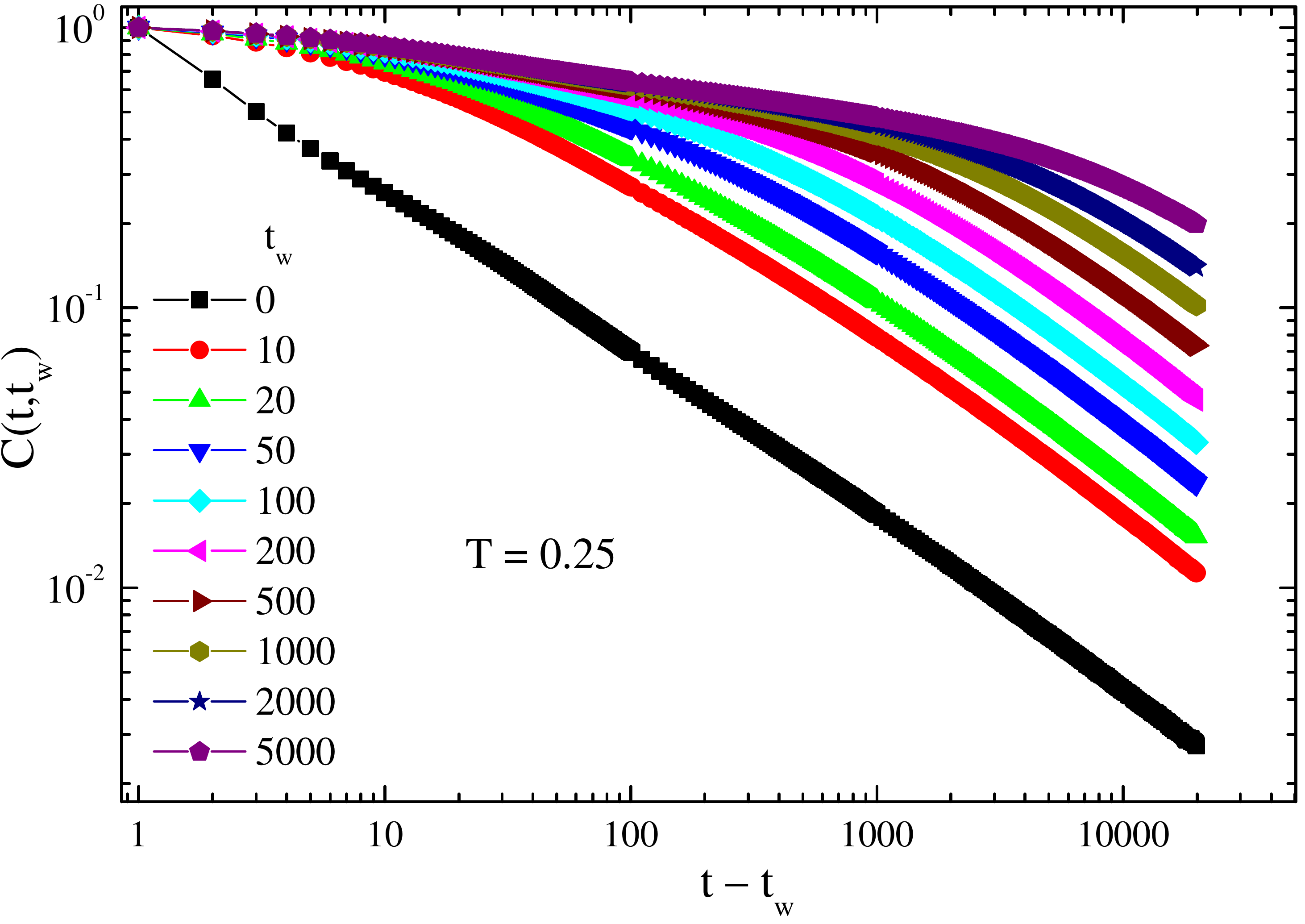}
\end{minipage}
\hfill
\vspace{0.3cm}
\begin{minipage}[t]{0.46\textwidth}
\includegraphics[width=0.9\linewidth]{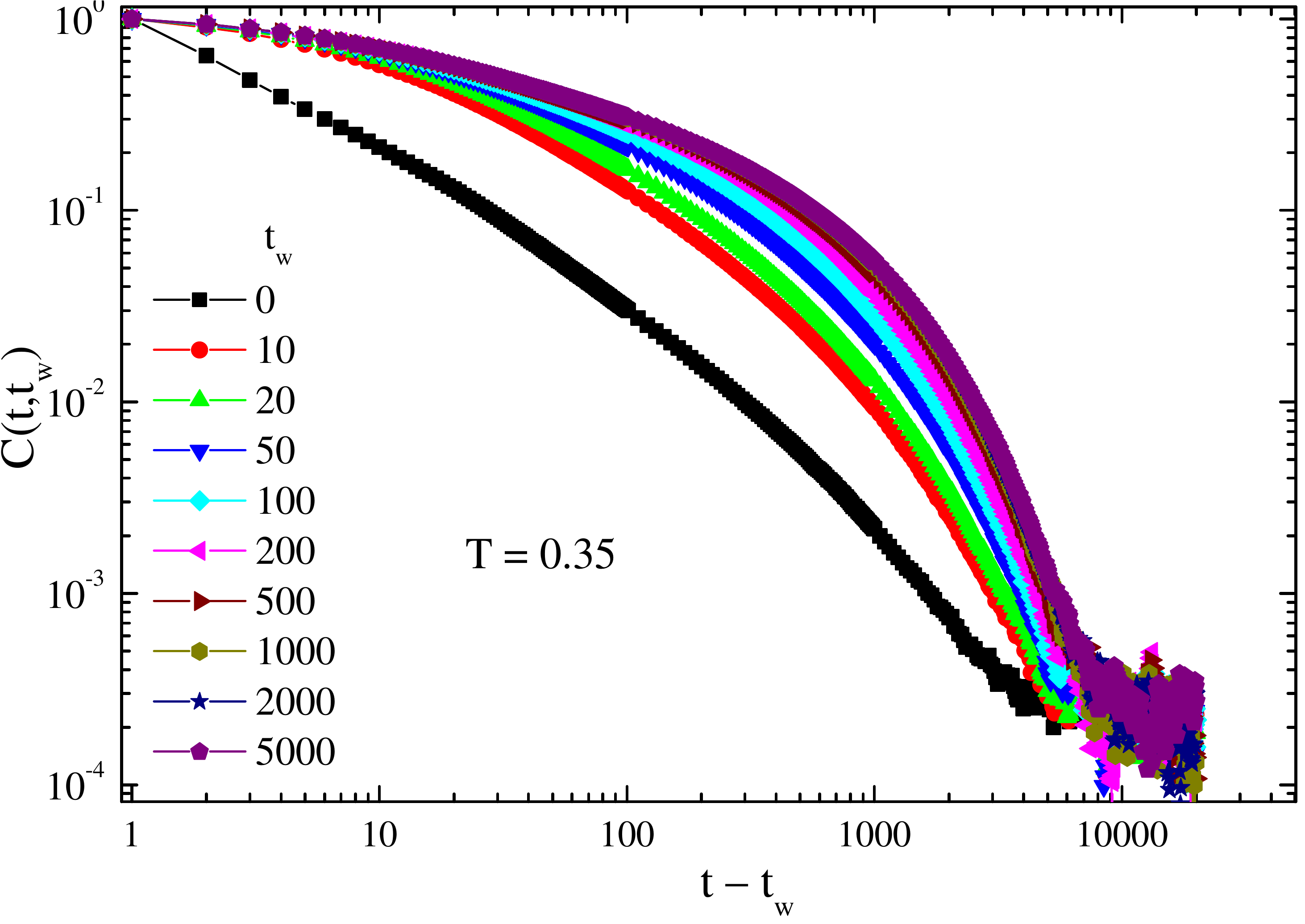}	
\end{minipage}
\vspace{-0.2cm}
\caption{\label{ACF_4} (Color online) Time dependence of autocorrelation spin function for high-temperature initial state, $L=240$, and $T=0.25$ (left plot) and $T=0.35$ (right plot).}
\label{Cttw}    
\end{figure*}

\begin{figure*}[t]
\begin{minipage}[h]{0.48\linewidth}
\center{	\includegraphics[width=0.9\textwidth]{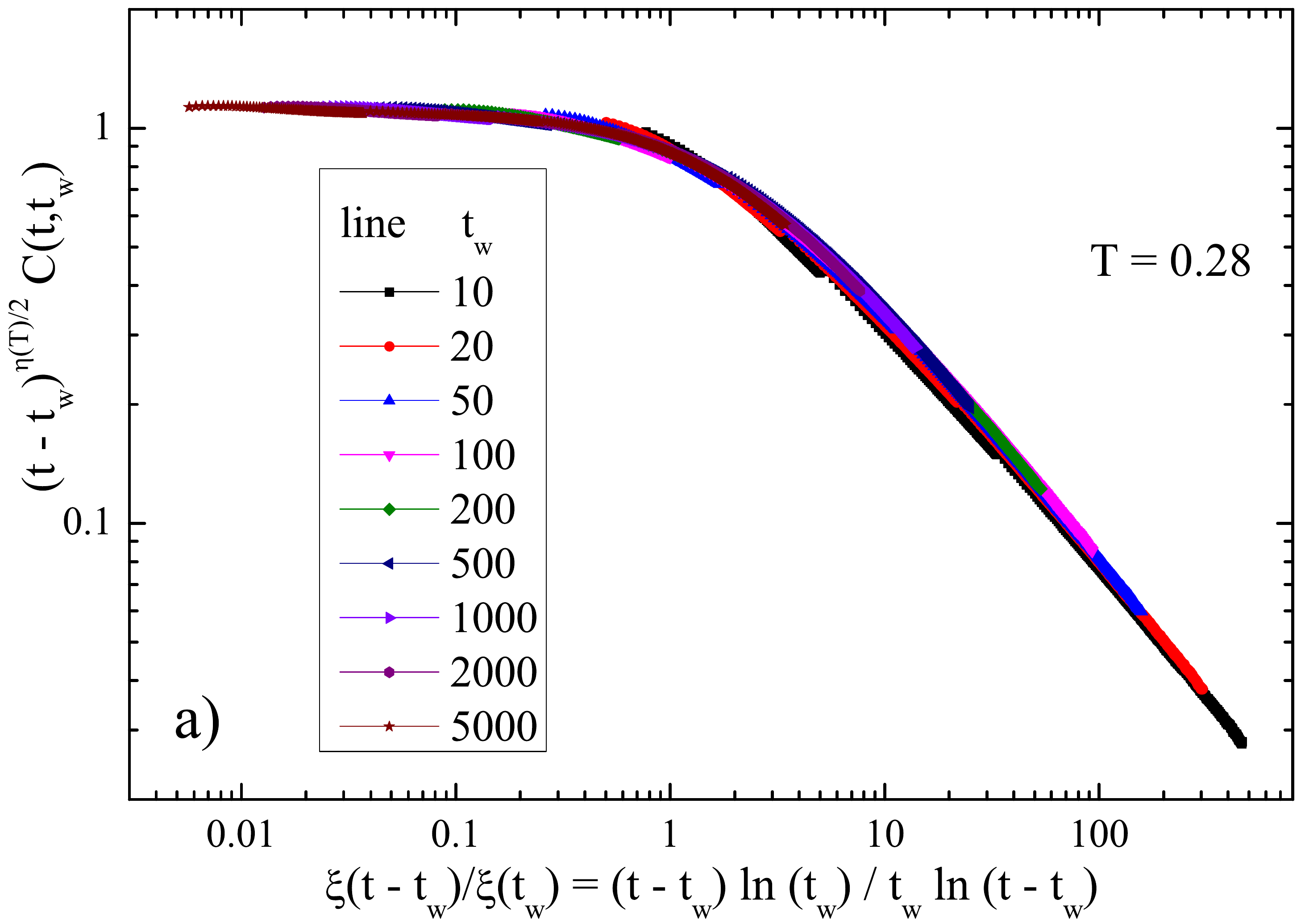}	
}
\end{minipage}
\hfill
\begin{minipage}[h]{0.48\linewidth}
\center{	\includegraphics[width=0.9\textwidth]{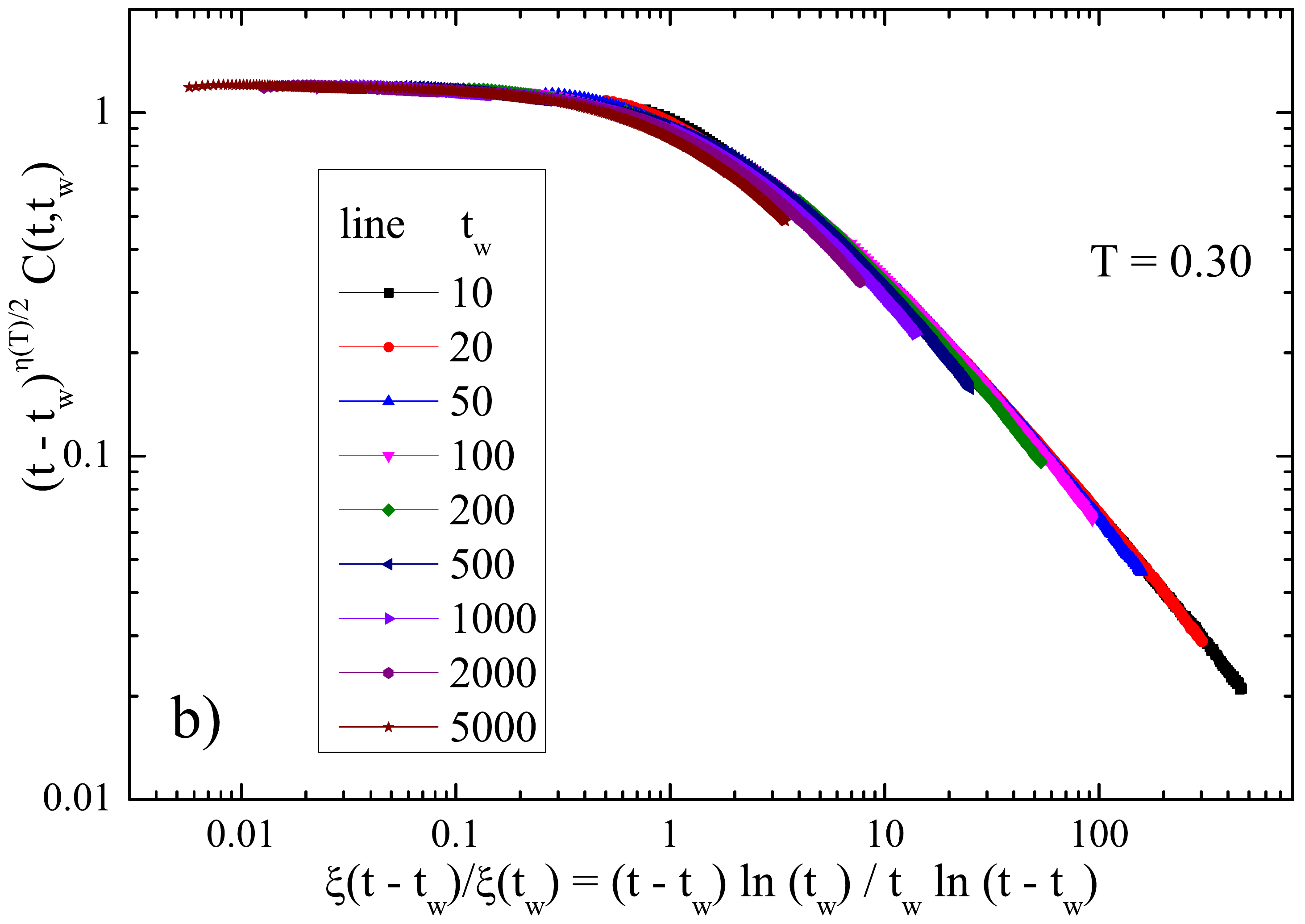}  
}
\end{minipage}
\vfill
\vspace{-0.0cm}
\begin{minipage}[h]{0.48\linewidth}
\center{    \includegraphics[width=0.9\textwidth]{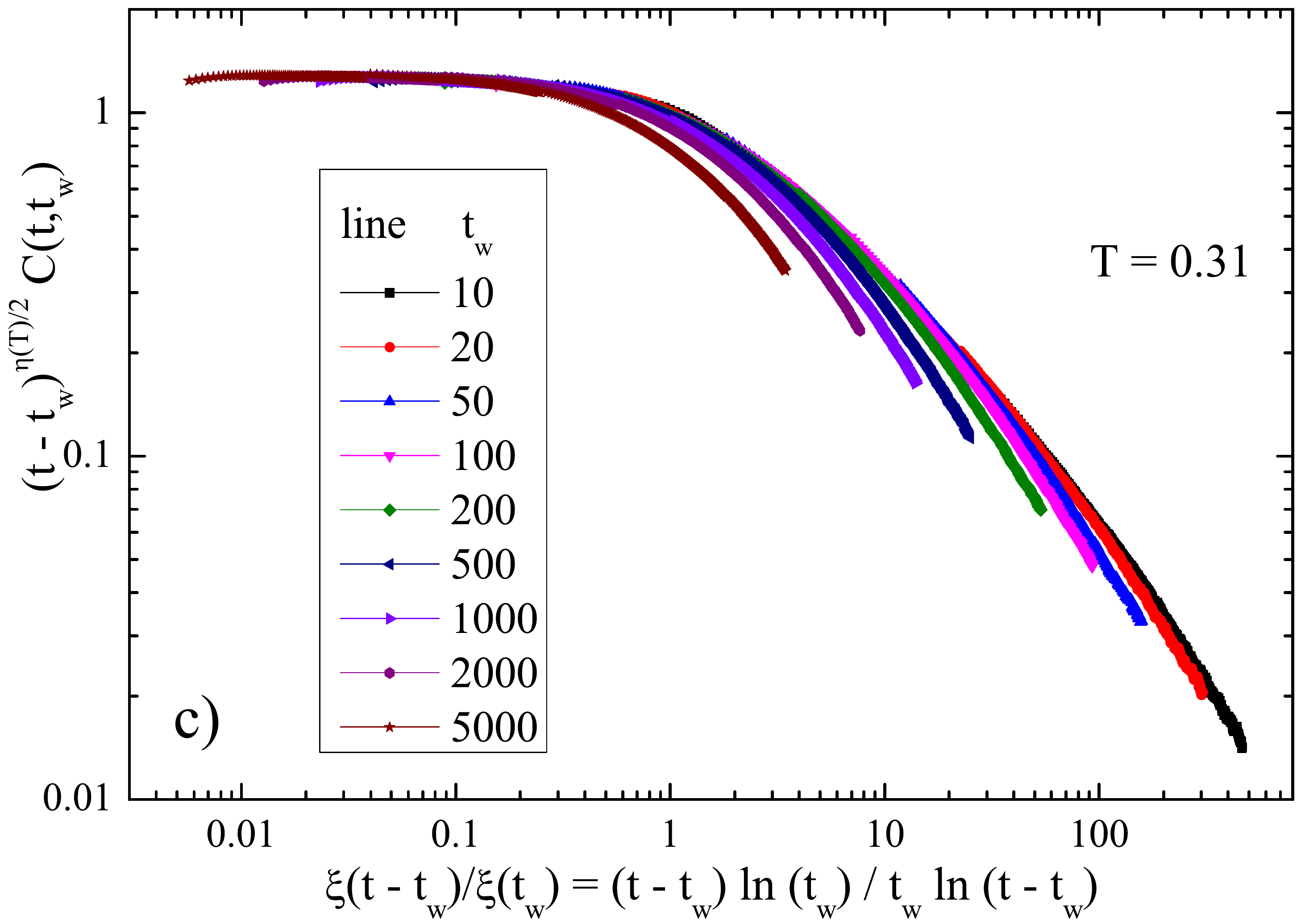}	
}
\end{minipage}
\hfill
\begin{minipage}[h]{0.48\linewidth}
\center{    	\includegraphics[width=0.92\textwidth]{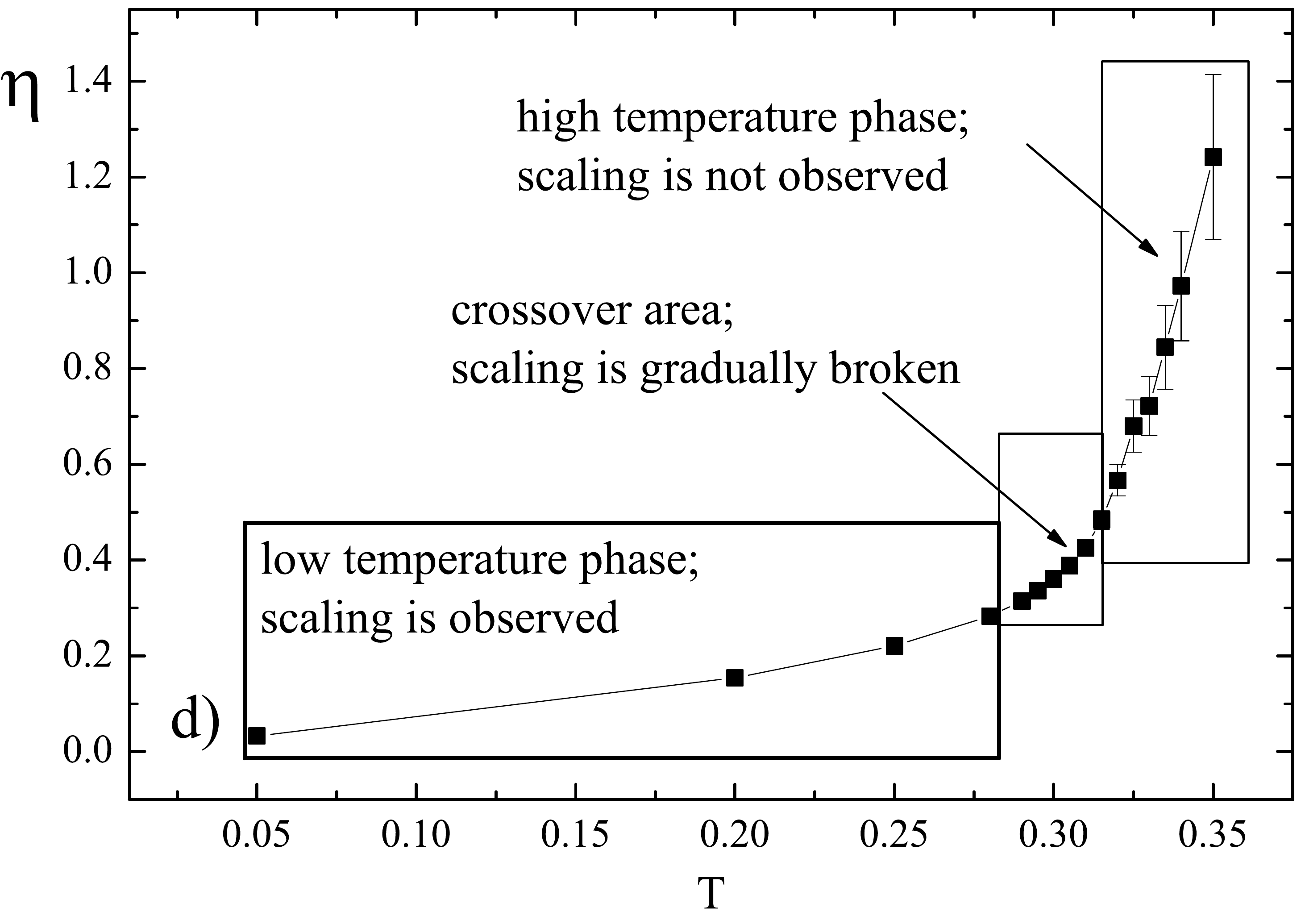}	
}
\end{minipage}
\caption{\label{Sc_ACF_1} (Color online)
(a-c) Scaling for the autocorrelation function of the system with $L=240$; $T=0.28$, $0.3$, and $0.31$ from the high-temperature initial state. (d) The temperature dependence of the anomalous dimension $\eta(T)$.}
\label{ris:experimentalcorrelationsignals}
\end{figure*}

The obtained time dependence of the correlation length can also be compared to that for the square lattice (see lower part of Fig. \ref{xsi}). In case of square lattice the $(t/\ln t)^{1/2}$ behavior is also obtained (although only in a rather restricted time range at low temperatures), presumably because of the presence of skyrmion quarks \cite{Fateev,Bukhvostov} (called also 'zindons' \cite{Diakonov}), which interact via logarithmic potential. Although these quarks are at the effective temperature $\approx 2 T_{\rm BKT}$, the correlation length appears to be exponentially large at low temperatures, providing possibility for approximate scaling. More detail discussion of the square lattice will be presented elsewhere.

\subsection{Spin autocorrelation function}

Let us now turn to the time dependence of the autocorrelation function 
\begin{equation} \label{ACF}
C(t,t_w)= \frac{1}{N} \sum\limits_{i}
\left<\mathbf{S}_i(t)\mathbf{S}_i(t_w)\right>.
\end{equation} 
We have observed presence of the aging effects (slowing of the relaxation processes by 
increasing the waiting time $t_w$) in the system for temperatures 
$T<0.35$ (see Fig. \ref{Cttw}); for higher temperatures obtained dependencies can not be attributed 
to the aging effects, they rather show non-uniform in time 
nonequilibrium relaxation. 
With the start from the initial high-temperature state we find slowdown in the time decay of the autocorrelation function with increasing waiting time $t_w$, while with the start from the initial low-temperature state the observed effects are opposite, and autocorrelation function decay is accelerated with increasing waiting time $t_w$.


The important feature of the aging is the scaling behavior, which  is obeyed by autocorrelation function in accordance with the general considerations of critical dynamics
\cite{Henkel,Henkel_SpinGlass},
\begin{equation}
	C(t,t_w)=(t-t_w)^{\eta/2} \Phi(\xi(t-t_w)/\xi(t_w)),
    \label{Scaling}
\end{equation}
where $\eta$ is the anomalous dimension, depending on the temperature.
On Fig.~\ref{Sc_ACF_1}a-c we present obtained scaling laws for two-time dependencies 
of the autocorrelation function according to the Eq. (\ref{Scaling}). 
One can see that in the low temperature phase the scaling of the autocorrelation function is fulfilled, but it begins to break  after passing to the high-temperature phase $T>\Tum$. On Fig. 
\ref{Sc_ACF_1}d we present the extracted temperature dependence of the critical exponent $\eta$. 
Surprisingly, at the vortex unbinding temperature $\Tum$ the critical exponent 
is close to the value $\eta(T_{\rm BKT})=0.25$ for the $XY$-model. 
We expect that critical exponent $\eta$ can be also observed in principle in static correlation function at the scales $a\ll r\ll \xi_{\rm sw}$. Note that in the considered system nonzero value of $\eta$ is generated by vortex pairs; in contrast to XY model spin-waves alone do not generate anomalous dimension \cite{Azaria_NPB_1993}.

\section{Conclusions}

In conclusion, we have calculated time dependence of the correlation length and spin correlation functions for Heisenberg antiferromagnet on the triangular lattice. 
From our consideration we conclude that on the intermediate time range the scaling properties of the frustrated Heisenberg model on the triangular lattice are very similar to those for the XY model: we observe $(t/\ln t)^{1/2}$ time dependence of the correlation length and fulfillment of the scaling law (\ref{Scaling}) in a broad time range at $T\le \Tum$, despite finite value of the equilibrium correlation length in this temperature range. This reflects vanishingly small equilibrium concentration of unbound vortices at low temperatures $T<\Tum$, and surprisingly weak effects of interaction between vortices and spin-wave degrees of freedom. 
Our results demonstrate also significant difference in the dynamics of the system above and below $\Tum$, such that the obtained temperature $\Tum$ can be considered as the temperature of the dynamic transition (or, at least, sharp crossover), related to the vortex unbinding.

Therefore, we have shown, that in the intermediate time range they fulfill scaling relations, identical to those, which were found earlier for the XY model. The dynamic properties of the triangular lattice antiferromagnets in this time range originate mainly from the $\mathbb{Z}_2$-vortices, which contribution to the dynamic properties is similar to that of the vortices of the XY model. At very short times we have observed some deviations from the scaling laws, which originate from the absence of well defined spin-wave and vortex excitations in the lack of short-range magnetic order. At very long times the contribution of the (equilibrated) spin waves, which provide finite correlation length below $\mathbb{Z}_2$-vortex unbinding transition/crossover temperature, becomes important.

{\it Acknowledgements}. 
The work was performed within the grant No. MD-6024.2016.2 of Russian Federation President (I.P., P.P.), themes  ``Quant'' 01201463332 (A. I.) and ``Electron'' 01201463326 (A. K.) of FASO, Russian Federation, and partially supported  within  the  projects of Russian Foundation of Basic Research 14-32-50632-mol-nr (I.P.), 14-02-00953a (A.I., A.K., I.P.) and 17-02-00279a (I.P., P.P). The calculations are performed on the Supercomputing Center of Lomonosov Moscow State University, Moscow Joint Supercomputer Center and St. Petersburg Supercomputer Center of the Russian Academy of Sciences.




\end{document}